\documentclass[twocolumn,superscriptaddress,
	showpacs,amsfonts,amssymb,amsmath,floats,aps]{revtex4-1}
\usepackage{graphicx}
\usepackage{dcolumn}
\usepackage{multirow}

\usepackage{braket}

\def\e{\varepsilon}

\def\w{\omega}

\def\k{\vec{k}}
\usepackage{upgreek}
\usepackage{bbold}

\usepackage{placeins}

\begin{document}
%----------------------------------

\title{Effective low-energy description of the two impurity Anderson model:\\
RKKY interaction and quantum criticality}

\author{Fabian Eickhoff}
\affiliation{Theoretische Physik 2, Technische Universit\"at Dortmund, 44221 Dortmund, Germany}
\author{Benedikt Lechtenberg}
\affiliation{Department of Physics, Kyoto University, Kyoto 606-8502, Japan}
\author{Frithjof B. Anders}
\affiliation{Theoretische Physik 2, Technische Universit\"at Dortmund, 44221 Dortmund, Germany}

\date{\today}

%----------------------------------
\begin{abstract}
%----------------------------------

We show that the  RKKY interaction  in the  two-impurity Anderson model
comprise two contributions: a ferromagnetic part stemming from the symmetrized hybridization functions and an anti-ferromagnetic part. 
We demonstrate that this anti-ferromagnetic contribution can also be generated 
by an effective local tunneling term between the two impurities.
This tunneling can be analytically calculated for particle-hole symmetric impurities.
Replacing the full hybridization functions by the symmetric part and this  
tunneling term leads to the identical low-temperature fixed point  spectrum in the 
numerical renormalization group.
Compensating this tunneling term is used to restore the  Varma-Jones quantum critical point 
between a strong coupling phase and a local singlet phase
even  in the absence of particle-hole symmetry in the hybridization functions.
We analytically investigate the spatial frequencies  of the effective tunneling term 
based on the combination of the 
band dispersion and the shape of the Fermi surface. Numerical renormalization group calculations  provide 
a comparison of the distance dependent tunneling term and the local
spin-spin correlation function. 
Derivations between the spatial dependency of the full spin-spin correlation function and
the textbook RKKY interaction are reported.

%----------------------------------
\end{abstract}
%----------------------------------

\maketitle
%----------------------------------

%----------------------------------
\section{Introduction}
%----------------------------------

Using technology based on  quantum-mechanical phenomena  for efficient computations, 
requires the realization of quantum bits, which might be 
implemented via quantum impurity systems \cite{lit:Loss_DiVincenzo1998,lit:Loss_DiVincenzo1999,lit:Trauzettel2007,lit:Greilich2006,lit:Glazov2013}.
Magnetic adatoms and molecules on surfaces 
as well as nano-structured gate controlled devices
could serve as smallest building blocks for such systems
\cite{lit:RevModPhys.76.323,lit:Spintronic_anisotropy_Misiorny2013,lit:Graphene_spintronics_Han2014,lit:Metals_on_silicon_Johll2014,lit:Defect_graphene_Yazyev2007,lit:Bork_Kroha2011,Esat2016,lit:spinfilter2010,lit:spintronic_review_Bogani2008,lit:Chemistry_Review_spintronics_Sanvito2011,lit:Review_Naber2007,lit:Organic_Dediu2009, lit:spin_diffusion_Drew2009,lit:Spinelli2015},
which allows to
combine the traditional electronics with novel spintronics and have been intensively studied in the last decades.

The two impurity Anderson model (TIAM) 
provides one of the simplest systems of two independent local moments
that indirectly couple through the conduction band of the host or substrate material.
It is particular interesting since it accounts for  the competition of 
two  mechanisms \cite{Jayaprakash1981,Jones1987,Jones1988,Sakai1990,Fye1987,Vojta2002,Sakai1993,Mitchell2015,Logan2011,Mitchell2011,Logan2012,Potthoff2012,Schwabe2012,Potthoff2013,Vojta2006,Oleg2018}
that influences the magnetic properties of the
ground state.
For a ferromagnetic Ruderman-Kittel-Kasuya-Yosida (RKKY) 
\cite{lit:RudermanKittel1954,lit:Kasuya1956,lit:Yosida1957} interaction $J_\text{RKKY}$, both impurity spins align
parallel and are screened by the itinerant conduction electrons,
while for strong antiferromagnetic interactions, both spins form a inter-impurity singlet which decouples from the conduction band. 

This observation triggered intensive research in the 1970s and 1980 in the context of Heavy Fermions \cite{Grewe91}
since it has been suggested that this competition provides a basic understanding of this class of materials: depending
on the interaction strength either a heavy Fermi liquid or a antiferrromagnetically order ground state is found driven
by the local singlet formation \cite{Doniach1977}. 
There is an ongoing discussion \cite{StockertSteglich2011,Kroha17}  
whether the change of ground states in a lattice system 
is connected to 
a quantum phase transition \cite{Jones1988,Jones1989,Jones1991} in a two-impurity system.

The transition between these two singlet phases in the TIAM 
is driven by
the ratio between the Kondo temperatur $T_\text{K}$ and $J_\text{RKKY}$ \cite{Silva1996,Zhu2011,Satoshi2006,Jones1991,Doniach1977,Ramires2016,Sakai1992,Jabben2012}. While a quantum critical point (QCP) separates both ground states in the presence
of a special kind of particle-hole (P-H)
symmetry  \cite{Jones1989,Jones1988,Sakai1992}, the quantum phase transition is replaced by a crossover if that symmetry
is broken \cite{Affleck1995}.
Including the  energy dependence in the impurity coupling function
generally leads to a P-H asymmetric model and, consequently, to a crossover behavior.
Therefore, the QCP found by Varma and Jones \cite{Jones1988,Jones1989,Jones1991} is a consequence of an 
oversimplification of the problem \cite{Fye1994}.
Recently, however,
it has been shown \cite{Lechtenberg2017} that for certain dispersions and distances between the impurities the
TIAM exhibits a QCP, separating two orthogonal ground states with different degeneracy.
This QCP is of different nature and has been experimentally observed in PTCDA-Au complexes on an Au surface 
\cite{Esat2016} and is driven by the additional direct tunneling term between the two neighboring molecular orbitals 
\cite{Esat2016,Lechtenberg2017}.

In this paper, we present an analytical, non-perturbative formula, based on a symmetry analysis of the parity dependent
and distance dependent hybridization function, which allows to map 
the emerging
scattering terms 
onto an effective tunneling $t^\text{eff}(\vec{R})$ between the impurities.
We present a full numerical renormalization group (NRG)
calculation \cite{Wilson1975,Krishna-murthy1980I,Krishna-murthy1980II,Bulla2008} to prove the equivalence of the effective and the original two impurity problem.

The construction of the effective tunneling term provides a new
insight to the nature of the
AF contribution to the RKKY interaction.
For the wide band limit we find that the AF contribution to $J_{\rm RKKY}$ is determined by $(t^\text{eff})^2/U$, 
where $U$ denotes the 
Coulomb interaction. This result is contrary to a separate two step transformation: (i) a Schrieffer Wolff transformation 
\cite{Schrieffer1966} onto the two impurity Kondo model
and (ii) the perturbative calculation of $J_{\rm RKKY}$ using this two impurity Kondo model, which would predict a $1/U^2$ dependency.
The distance dependence of the effective tunneling term can explain our numerical findings, that the impurity spin-spin correlation function
decays remarkably  slower than the textbook expression of the RKKY interaction, even for
a finite bandwidth of the conduction band.
We study the spatial anisotropy of this tunneling term on a simple cubic lattice and
find a surprising direct connection between slow (fast) spatial oscillations and
particle (hole) doping, that is beyond the standard $2k_\text{F}$ oscillations.

The understanding of the effective tunneling term enables us to
engineer the recovery of the Varma and
Jones quantum critical point in the TIAM for arbitrary distances,
even for a particle-hole symmetry broken model that
generically only shows a continuous change of the conduction electron
scattering phase.
Such additional local tunneling term can also  naturally occur  in neighboring molecular orbitals 
as shown by density functional theory \cite{Esat2016}.

This paper is structured as follows.
We start by defining the model, its mapping onto the parity eigenbasis
and the  fixed point (FP) structure of the NRG level flow in Sec.\ \ref{sec:theory} .
In Sec.\ \ref{sec:restoring_the_QCP}, we review the  
different types of P-H symmetries and derive the effective low-temperature description of the model,
based on an additional spatial-dependent local tunneling term $t^{\rm eff}(\vec{R})$.
This approach is applied in Sec.\ \ref{sec:results} to restore the QCP of  Varma and Jones by investigating the impurity
spectral functions, the scattering phase of the Green function and the NRG level flow.
We also cover the finite distance and finite bandwidth corrections to the impurity spin-spin correlation function.
Section \ref{sec:Sclattice} is devoted to the analytical analysis of the spatial frequencies governing the
spatial anisotropy of the spin-spin correlation function in a simple cubic lattice as function of the chemical potential, and,
therefore, the shape of the Fermi surface. We close with a summary in Sec.\ \ref{sec:conclusion}.
%

%----------------------------------
\section{Theory}
%----------------------------------
\label{sec:theory}

%----------------------------------
\subsection{Two Impurity Anderson Model}
%----------------------------------
\label{TIAM}

The Hamiltonian of the TIAM can be divided into three parts 
\begin{align}
\label{eqn:htiam}
H_{\text{TIAM}} &= H_{\text{imp}} + H_{\text{host}} + H_{\text{hyb}}.
\end{align}
The impurity part is given by
\begin{align}
\label{eqn:himp}
H_{\text{imp}} &= \sum_{l \in \{1,2\},\sigma}\epsilon^f_{l} f^{\dagger}_{l,\sigma}f_{l,\sigma}+
\frac{t}{2}\sum_{l,\sigma} f^{\dagger}_{l,\sigma}f_{\bar{l},\sigma}\nonumber\\
&+\frac{1}{2}\sum_{l\in\{1,2\},\sigma}U_l f^{\dagger}_{l,\sigma}f_{l, \sigma}f^\dagger_{l,\bar{\sigma}}f_{l,\bar{\sigma}}.
\end{align}
The operator $f^{(\dagger)}_{l,\sigma}$
destroys (creates) an electron with spin $\sigma=\pm$ on impurity $l$, whose onsite energy is labeled
by $\epsilon^f_l$.
$U$ denotes the onsite Coulomb repulsion. Furthermore, 
we also allow for a tunneling term $t$  between both impurities. 
Such an hopping term is realized in a system where the local orbitals are given by the lowest unoccupied molecular
orbitals of two neighboring molecule complexes that start to overlap at short distance and form dimers \cite{Esat2016}.

The metallic host is described by a free conduction band
\begin{align}
\label{eqn:hhost}
H_\text{host}&=\sum_{\vec{k},\sigma}\epsilon^{c}_{\vec{k}} c^{\dagger}_{\vec{k}, \sigma}c_{\vec{k}, \sigma},
\end{align}
where $c_{\vec{k}} (c^{\dagger}_{\vec{k}})$ is the annhilation (creation) operator of an
electron in the conduction band with dispersion $\epsilon^c_{\vec{k}}$. 
The interaction between the impurities and the host accounted for by
\begin{align}
\label{eqn:hhyb}
H_\text{hyb}&=\sum_{\substack{l \in \{1,2\}\\ \vec{k},\sigma}}\left(V_{l\vec{k} }c^{\dagger}_{\vec{k},\sigma}e^{i\vec{k}\vec{R}_{l}}f_{l,\sigma}+\text{h.c.}\right).
\end{align}
Here $V_{l\vec{k}}$ denotes the hybridization of the impurity located at position 
$\vec{R}_{l}$ with the conduction band state $\vec{k}$.

In the following, we consider the parity symmetric case, $V_{1\vec{k}}=V_{2\vec{k}} = V_{\vec{k}},
\e^f_0=\epsilon^f_1=\epsilon^f_2;\, U_1=U_2$, unless stated otherwise. 
Close to integer valence of one electron per impurity, 
a local moment is formed at intermediate temperatures \cite{Schrieffer1966} that is screened for $T\to 0$ 
\cite{Krishna-murthy1980I,Jayaprakash1981,Jones1987}. This is the case for $\e^f_0 \approx -U/2<0$ and will be the 
main focus
of this paper.

The hybridization induces an effective Heisenberg exchange interaction, the RKKY interaction, between the two impurities
that alters in sign with the characteristic spatial dependency of $\cos(2k_F R)/R^d$ for $Rk_\text{F}
\gg 1$ with $d$ being the
spatial dimension of the host, assuming a simplified energy dispersion of the conduction band.

We can also artificially add an additional direct Heisenberg exchange interaction $J_{12}\vec{S}_1\vec{S}_2$ to the full
two impurity Hamiltonian 
\begin{eqnarray}
H'_\text{TIAM}(J_{12})&=& H_\text{TIAM} + J_{12}\vec{S}_1\vec{S}_2
\end{eqnarray}
as an external control parameter for the investigation of the QCP of the Varma-Jones (VJ) type.

\subsection{Energy representation of the TIAM Hamiltonian}

It is convenient
to convert to a site-dependent and energy-dependent operator
\cite{Wilson1975,Affleck1995}, defining
\begin{eqnarray}
c_{l,\sigma}(\e) &=& \sqrt{\frac{\pi}{\Gamma(\epsilon)}} \sum_{\vec{k}} V_{\vec{k}}
\delta(\epsilon-\epsilon_{\vec{k}}^c) e^{i\vec{k}\vec{R}_l}c_{\vec{k},\sigma}
\end{eqnarray}
where the hybridization function $\Gamma(\epsilon)$,
\begin{eqnarray}
\Gamma(\epsilon) &=& \pi \sum_{\vec{k}} |V_{\vec{k}}|^2
\delta(\epsilon-\epsilon_{\vec{k}}^c) 
\,
\end{eqnarray}
is determined from the equal site anti-commutator $\{c_{l,\sigma}(\e),c^\dagger_{l,\sigma'}(\e') \} = \delta_{\sigma\sigma'}\delta(\e-\e')$.

The  hybridization takes the form
\begin{eqnarray}
H_\text{hyb}&=&\sum_{l \in \{1,2\}} \int_{-D}^{D} d\e \sqrt{\frac{\Gamma(\epsilon)}{\pi}} f^\dagger_{l,\sigma}c_{l,\sigma}(\e) + h.c
\end{eqnarray}
in the continuum limit, with the bandwidth $2D$ of the host. 

Using the effective hybridization matrix element $V$ given by
\begin{eqnarray}
\label{eqn:V-l-def}
\int_{-D}^{D} d\e \Gamma(\epsilon)  &=& V^2 \pi
\end{eqnarray}
we can define an effective conduction band density of states $\rho(\e) = \Gamma(\e)/(\pi V^2)$.

The operators $c_{l,\sigma}(\e)$ are connected to the same single conduction band and are
not linear independent. 
Therefore, they are combined to
parity eigenstates \cite{Jayaprakash1981,Jones1988,Jones1989,Affleck1995,Esat2016,Lechtenberg2014,Borda2007}
with even (e) and odd (o) parity that anti-commute by symmetry.
The spatial dependence in this even-odd parity basis is included into the new 
orthogonal energy-dependent field operators
\begin{eqnarray}
c_{\mu,\sigma}(\e) &=&  \sqrt{\frac{\pi}{\Gamma_{\mu}(\epsilon)}} \sum_{\vec{k}} V_{\vec{k}} \delta(\epsilon-\epsilon_{\vec{k}}^c)
\left(e^{i\frac{\vec{k}\vec{R}}{2}}+ s_\mu e^{-i\frac{\vec{k}\vec{R}}{2}}\right)c_{\vec{k},\sigma},
\nonumber
\\
\end{eqnarray}
with $\mu\in\{e,o\}$, $\vec{R}= \vec{R}_1-\vec{R}_2$ and $s_e=1$, $s_o=-1$. The effective hybridization functions,
\begin{eqnarray}
\label{eq:def-gamma-mu}
\Gamma_e(\epsilon,\vec{R})&=&
\pi \sum_{\vec{k}} |V_{\vec{k}}|^2\delta(\epsilon-\epsilon_{\vec{k}}^c)\, \text{cos}^2(\vec{k}\vec{R}/2),\nonumber\\
\Gamma_o(\epsilon,\vec{R})&=&
\pi \sum_{\vec{k}}  |V_{\vec{k}}|^2\delta(\epsilon-\epsilon_{\vec{k}}^c)\, \text{sin}^2(\vec{k}\vec{R}/2),
\end{eqnarray}
are defined such that the standard anti-commutation relation
$\{c_{\mu,\sigma}(\e),c^\dagger_{\mu^\prime,\sigma^\prime}\}(\e^\prime)=\delta(\epsilon-\epsilon^\prime)\delta_{\mu,\mu^\prime}
\delta_{\sigma,\sigma^\prime}$ is fulfilled. They determine the effective coupling of the two different flavors even and odd to the
impurity and obey
\begin{eqnarray}
\Gamma(\e) &=& \Gamma_e(\e,\vec{R})+\Gamma_o(\e,\vec{R}) \, .
\end{eqnarray}

Introducing a even/odd parity basis also for the impurity operators,
\begin{eqnarray}
f_{\mu,\sigma}&=&\frac{1}{\sqrt{2}}\left( f_{1,\sigma}+ s_\mu f_{2,\sigma}\right)
\, ,
\end{eqnarray}
yields a flavor diagonal hybridization between the impurities and these  even/odd conduction bands
\begin{eqnarray}
\label{eqn:hhybevenodd}
H_\text{hyb}&=
& 
\sum_{\mu\in\{e,o\},\sigma}
\int_{-D}^{D}d\epsilon
\sqrt{\frac{\Gamma_\mu(\e,\vec{R})}{2\pi}}
c^\dagger_{\mu,\sigma}(\e) f_{\mu,\sigma}+\text{h.c.}\,.
\nonumber \\
\end{eqnarray}

By extracting the effective flavor coupling constant $V_\mu$,
\begin{eqnarray}
V^2_\mu(\vec{R}) \pi &=& \int_{-D}^{D} d\e \Gamma_{\mu}(\epsilon,\vec{R})  
\end{eqnarray}
we define the effective density of states of the flavor bands by normalization \cite{BullaPruschkeHewson1997,Bulla2008}
\begin{eqnarray}
\label{eqn:bar-rho-mu}
\bar \rho_{\mu}(\e,\vec{R}) &=&  \frac{1}{V^2_\mu(\vec{R}) \pi } \Gamma_{\mu}(\epsilon,\vec{R}) .
\end{eqnarray}
One can always find a proper normalized $\bar \rho_{o}(\e,\vec{R})$ 
in the limit $\vec{R}\to 0$: The decoupling of the odd
conduction band is accounted for by $V_o\to 0$. 
To this end, the hybridization can be expressed as
\begin{eqnarray}
\label{eqn:hhybevenodd-new}
H_\text{hyb}&=
& 
\sum_{\mu\sigma} V_\mu(\vec{R})
\int_{-D}^{D}d\epsilon
\sqrt{\bar \rho_{\mu}(\e,\vec{R})}
c^\dagger_{\mu,\sigma}(\e) f_{\mu,\sigma}+\text{h.c.}\,,
\nonumber \
\end{eqnarray}
separating the coupling strength to the impurity from the energy dependency of a normalized conduction band used to
construct the semi-infinite Wilson chains \cite{Jayaprakash1981,Jones1988,Jones1989,Affleck1995,Lechtenberg2014,Borda2007,Bulla2008}.
Note that the energy dependence of $\bar{\rho}_\mu(\e)$ generally destroys P-H symmetry.

\subsubsection{Low temperature fixed points}

In this section we briefly review the known low-temperature FP  structure of the TIAM model\cite{Jayaprakash1981,Jones1988,Jones1989,Affleck1995,Bulla2008}.
Since we are interested in the competition between the Kondo effect and the singlet formation due to the RKKY interaction,
we focus on the regime of singly occupancy of each impurity orbital.

Starting from a single impurity Anderson model in a parameter regime where the Schrieffer-Wolff transformation \cite{Schrieffer1966}
is applicable, the low-temperature FP is given by a strong-coupling (SC) FP describing the Kondo effect. The crossover to this
FP  is governed by a  non-analytic energy scale $T_\text{K}$ that is exponentially small in terms of the bar coupling constants. 
The local spin of the magnetic impurity is dynamically screened by the 
conduction electrons and the remaining conduction electron degrees of freedom decouple from the impurity.
Thus the  SC FP agrees with that of a free electron gas (FEG) with one electron removed that forms the Kondo singlet.
The conduction electrons  close to the Fermi energy acquire a  phase shift of $\delta$  in accordance
with Friedel's sum rule \cite{Langreth1966,Anders1991}.

While P-H symmetry pins the phase shift to $\delta=\pi/2$, P-H asymmetry leads to potential
scattering in the conduction band which changes the phase shift continuously.
The SC FP is given \cite{Krishna-murthy1980I,Krishna-murthy1980II} by a P-H symmetric term $H^{SC}_{PH}$
\begin{eqnarray}
\label{eq:H-SC-FP}
H^{SC}(K) &=& H^{SC}_{PH} + K\sum_\sigma\left(\bar c_{0\sigma}^\dagger \bar c_{0\sigma} -1\right)
\end{eqnarray}
and a marginal operator breaking P-H symmetry that is parameterized by the constant $K$. The operators $\bar c_{n\sigma}$ annihilate 
an electron with spin $\sigma$ on site $n=0,1,\cdots$ of the semi-infinite Wilson chain. All other scattering terms are irrelevant.
Below, we will make use of the fact that the FP is full characterized by a single constant $K$ defining a line of renormalization group (RP) FPs  \cite{Krishna-murthy1980I,Krishna-murthy1980II}.

In order to understand the low-temperature FP of the TIAM, we start from the Varma-Jones approximation who replaced
the energy dependent effective DOS by a constant,  $\bar \rho_{\mu}(\e)$ to $\rho_0$, enforcing P-H symmetry of the model. Since
only a FM RKKY interaction is dynamically generated by this simplification, an artificial local spin-spin coupling $J_{12}\vec{S}_1\vec{S}_2$
has been added. For a large FM $(-J_{12})\gg 0$, 
the locally favored triplet is screened by both conduction electron flavors by a two
stage Kondo effect since $V_o\not = V_e$. The FP is given by $H^{SC}_{PH}$ and $\delta_\mu=\pi/2$. For a large anti-ferromagnetic (AF) coupling,
$J_{12}\gg 0$, a local singlet is favored and the RG FP is given by those of a free electron gas $H^{FEG}_{PH}$ and  $\delta_\mu=0$. Since
P-H symmetry is only compatible with these two scattering phases, there must be a critical AF coupling at which the SC Kondo phase is replaced by the local singlet phase \cite{Affleck1995}.
This quantum critical point (QCP) occurs at $J_{12}/T_K\approx 2.2$ \cite{Jones1988,Jones1989}.\\
Once the full energy dependency of $\bar \rho_{\mu}(\e)$ required for the correct description of the RKKY interaction is taken into account,
the Varma-Jones QCP is replaced by a smooth crossover \cite{Silva1996}.

\section{Derivation of the effective tunneling term}
\label{sec:restoring_the_QCP}

Now we derive an analytical counter term to the bare Hamiltonian that allows to
restore the Varma-Jones QCP for arbitrary impurity distances.  The naive strategy would be to
add a suitable potential scattering term to the conduction electrons to restore P-H symmetry at the FP \cite{LechtenbergThesis2016}. 
The parameter, however $\bar K_\mu$ is subject to an RG flow,
 and it is very cumbersome to iteratively determine $\bar K_\mu$. In addition, the physical insight gained from such a
term is limited.

It turns out that modifying the tunneling term in $H_{\text{imp}}$ defined in Eq.\ \eqref{eqn:himp} has the identical effect and
the required $t^{\rm eff}$ can be analytically derived from the coupling functions $\Gamma_\mu(\e,\vec{R})$.

There are essential two  scenarios. 
(i) If the impurities are P-H symmetric ($\epsilon_l=-U/2$), 
there is a strong symmetry restriction of the type of potential scattering counter term. 
In this case
the low-temperature FP becomes P-H symmetric and $\delta_e=\delta_o=\pi/2$.
(ii) Local P-H asymmetry on the impurities
generates potential scattering in at least one of the channels.
Although we can modify these scattering terms to achieve $\delta_e=\delta_o$, 
which is sufficient to restore the Varma-Jones QCP \cite{Affleck1995,Zhu2006},
the scattering phases differ from $\pi/2$.

Since the parameters necessary to restore the QCP  
can be analytical derived only for first scenario, we start with
$\epsilon_l=-U/2$ and come  back to the second case latter.

\subsection{Particle-Hole symmetry and potential scattering}
\label{sec:phs}

We now review the connection between P-H symmetry and the arising potential scattering terms
discussed by  Affleck et al.\ \cite{Affleck1995}.

The TIAM with a P-H symmetric impurity can exhibit two different types of particle-hole symmetries.
The first type of P-H transformation requires a flavor diagonal transformation
\begin{align}
\label{eqn:phs1}
c_{\mu,\sigma}(\e)&\rightarrow c^\dagger_{\mu,\sigma}(-\e),
\end{align}
and is a symmetry of the Hamiltonian if the effective conduction bands are compatible with
\begin{align}
\label{eqn:phs1Nevenodd}
\bar \rho_\mu(-\epsilon,\vec{R})&=\bar \rho_\mu(\epsilon,\vec{R}).
\end{align}

However, the system can also be invariant under the second, flavor exchanging
P-H transformation
\begin{align}
\label{eqn:phs2}
c_{e/o,\sigma}(\e)&\rightarrow c^\dagger_{o/e,\sigma}(-\e)
\end{align}
if $V^2_\mu(\vec{R})\bar \rho_\mu(-\epsilon,\vec{R})$ satisfy the relations
\begin{align}
\label{eqn:phs2Nevenodd}
V^2_e(\vec{R})\bar \rho_e(-\epsilon,\vec{R})&=V^2_o(\vec{R})\bar \rho_o(\epsilon,\vec{R}).
\end{align}

In general the potential scattering terms generated in higher order of perturbation theory take the form
\begin{align}
\label{eqn:hscatt}
H_\text{s}=\sum_{\mu\in\{e,o\}}\int_{-D}^D \text{d}\epsilon \text{d}\epsilon^\prime\left[S_\mu(\epsilon,\epsilon^\prime)c^\dagger_{\mu}(\e)c_{\mu}(\e^\prime)
\right].
\end{align}
If the original problem is P-H symmetric, the effective potential scattering term must also  satisfy the special type of
symmetry transformation. Depending on the type of P-H symmetry, we require
\begin{align}
&\text{first type}\quad\,\longrightarrow\quad S_{e/o}(\epsilon,\epsilon^\prime)=-S_{e/o}(-\epsilon,-\epsilon^\prime),\nonumber\\
&\text{second type}\longrightarrow\quad S_{e/o}(\epsilon,\epsilon^\prime)=-S_{o/e}(-\epsilon,-\epsilon^\prime),
\end{align}
thus the scattering function must vanish at  zero-energy in the presence of the first type of symmetry, 
whereas the second type only requires a connection
between the even and odd channels:
\begin{align}
\label{eqn:phaseshift}
&\text{first type}\quad\,\longrightarrow\quad S_{e/o}(0,0)=0,\nonumber\\
&\text{second type}\longrightarrow\quad S_{e/o}(0,0)=-S_{o/e}(0,0).
\end{align}
Since the zero-energy scattering terms in the even and odd channels in general lead to different phase shifts $\delta_{e/o}$
and hence destroy the QCP, 
only the first type of P-H symmetry ensures the existence of a QCP in the TIAM automatically.

\subsection{Low energy description and effective tunneling}

Even for a P-H symmetric dispersion $\epsilon^{c}_{\vec{k}}$ 
of the original problem, the effective densities of states $\bar \rho_{\mu}(\e,\vec{R})$ defined in 
 Eq.\ \eqref{eqn:bar-rho-mu} will generally not comply with any of the two types of P-H symmetries.
However, one can divide $\bar \rho_{\mu}(\e,\vec{R})$ into the two contributions
\begin{eqnarray}
\label{eq:bar-rho-pm}
\bar \rho_{\mu}^{(\pm)}(\e,\vec{R}) &=& \frac{1}{2}
\left[ \bar \rho_{\mu}(\e,\vec{R}) \pm \bar \rho_{\mu}(-\e,\vec{R})\right],
\end{eqnarray}
While $\bar \rho_{\mu}^{(+)}(\e,\vec{R})$ satisfies Eq.\;\eqref{eqn:phs1Nevenodd} and is normalized.

$\bar \rho_{\mu}^{(-)}(\e,\vec{R})$  has a vanishing integral spectral weight and, therefore,
cannot be interpreted as an effective bath. This term breaks the P-H symmetry of
first type and contributes to the scattering terms.

Consequently, the Hamiltonian of each  conduction band flavor $\mu$ can be decomposed into
\begin{eqnarray}
H_{\text{host},\mu}&=& H_{\text{host},\mu}^+ + \Delta H_{\text{host},\mu}^-\,,
\end{eqnarray}
where $H_{\text{host},\mu}^+$ describes an fictitious bath with P-H symmetry of the first type,
while $\Delta H_{\text{host},\mu}^-$ stems from redistribution of spectral weight due to $\bar \rho_{\mu}^{(-)}(\e,\vec{R})$
that can be accounted for by an appropriately chosen scattering function $S_\mu(\epsilon,\epsilon^\prime)$
in Eq.\ \eqref{eqn:hscatt}.

We make use of the fact \cite{Krishna-murthy1980I,Krishna-murthy1980II,LechtenbergThesis2016} that
the P-H symmetry  breaking  leads to a modification of the fixed point Hamiltonian controlled by a single scattering parameter $K_\mu$
in each band, such that we alternatively can approximate the host by
\begin{eqnarray}
\label{eqn:heff}
 H_{\text{host},\mu}&\approx& H_{\text{host},\mu}^+ + K_\mu
 \sum_\sigma\left(\bar c_{0\mu\sigma}^\dagger \bar c_{0\mu \sigma} -1\right)
\,.
\end{eqnarray}
If $\rho(\epsilon)$ as defined below Eq.\ \eqref{eqn:V-l-def} is invariant under  energy inversion, i.~e.\ $\rho(\epsilon)=\rho(-\epsilon)$,
one can show 
that $V^2_e(\vec{R})\bar \rho_{e}^{(-)}(-\e,\vec{R}) = -V^2_o(\vec{R})\bar \rho_{o}^{(-)}(\e,\vec{R})$. As a consequence
$K_e= -K_o$ or $K_\mu = s_\mu K$, and the problem is reduced to a single parameter that determines the low-temperature effect
of $\bar \rho_{e}^{(-)}(\e,\vec{R})$.

Now we turn to the full Hamiltonian of the  TIAM that also contains the
local impurity degrees of freedom and the coupling between
both subsystems. An impurity interaction that is invariant under the transformation
\begin{align}
\label{eqn:phs2imp}
f_{e/o,\sigma}\rightarrow f^\dagger_{o/e,\sigma}\mspace{60mu}\nonumber\\
\Leftrightarrow \quad f_{1,\sigma}\rightarrow f^\dagger_{1,\sigma}\,;\quad f_{2,\sigma}\rightarrow -f^\dagger_{2,\sigma},
\end{align}
\noindent but not under the transformation
\begin{align}
\label{eqn:phs1imp}
&f_{e/o,\sigma}\rightarrow f^\dagger_{e/o,\sigma}\nonumber\\
\Leftrightarrow \quad &f_{1/2,\sigma}\rightarrow f^\dagger_{1/2,\sigma},
\end{align}
is only compatible with the second type of P-H transformation and hence inevitably generates potential scattering terms
in the form of $K_e=-K_o\not=0$ in the low-energy FP that is compatible to Eq.\;\eqref{eqn:phaseshift}.
Therefore we can replace the scattering 
terms in Eq.\;\eqref{eqn:heff} by an effective impurity interaction $H^\text{eff}_\text{imp}$ 
that leads to the same low-energy FP.
Note that the invariance of $H^\text{eff}_\text{imp}$ under the transformations of Eqs.\;\eqref{eqn:phs1imp} 
ensures that the full Hamiltonian in Eq.\;\eqref{eqn:heff} remains P-H asymmetric.

The only parity-conserving single particle term involving only impurity degrees of freedoms
that is invariant under local P-H transformation of the second type, Eq.\ \eqref{eqn:phs2imp}, but not under \eqref{eqn:phs1imp} is given by
\begin{align}
\label{eqn:eff}
\mathrm{H_\text{imp}^\text{eff}}&=\frac{t^\text{eff}}{2}\sum_{\sigma}\left(f^\dagger_{e,\sigma}f_{e,\sigma}-f^\dagger_{o,\sigma}f_{o,\sigma}\right)\nonumber\\
&=\frac{t^\text{eff}}{2}\sum_{\sigma}\left(f^\dagger_{1,\sigma}f_{2,\sigma}+f^\dagger_{2,\sigma}f_{1,\sigma}\right).
\end{align}
This term is parameterized by a single parameter $t^\text{eff}$ that has a simple physical interpretation:
It describes an additional electron tunneling term between the two impurities and is fully compatible with $H_{\text{imp}}$.
Mahmoud et al. already mentioned the existence of such an effective charge exchange in the non-interacting two impurity Anderson model  on a lattice \cite{Mahmoud2017}.

\subsection{Estimate of effective tunneling}

One of the key messages of this paper is that one can subtract an appropriately chosen 
local impurity counter term $\mathrm{H_\text{imp}^\text{eff}} $ 
in order to restore  the Varma-Jones QCP. It is well established \cite{Affleck1995,Silva1996} that the
QCP is destroyed only by scattering terms compatible with the P-H symmetry of the second kind, leading to different scattering phases
in the even and the odd channel. The goal of the counter term is to produce  identical scattering phase $\delta_e=\delta_o$ for $T,\w\to 0$.

In order to gain some insight and 
actually calculate $t^\text{eff}$ in a certain limit, we demand that the low-temperature FP of the  full model $H_\text{TIAM}$ 
augmented with a counter term $H_\text{imp}^\text{eff}$
\begin{eqnarray}
\label{eqn:Heff}
H_\text{TIAM}^{\rm eff} &=&  H_\text{TIAM} -H_\text{imp}^\text{eff}
 \end{eqnarray}
is identical to those of the effective model  $H^+_\text{TIAM}\overset{!}{=} H^{\text{eff}}_\text{TIAM}$
where the full DOS $\bar \rho_{\mu}(\e,\vec{R})$ 
has been replaced by $\bar \rho^{(+)}_{\mu}(\e,\vec{R})$ of
Eq.\ \eqref{eqn:hhybevenodd-new}.
If the parameters of the impurities are P-H symmetric, i.\ e. $\e_0^f+ U/2=0$, 
the scattering phases of $H^+_\text{TIAM}$ 
is distance independent and equal $\delta_e=\delta_o= \pi/2$ 
and likewise in $H_\text{TIAM}^{\rm eff}$.

\begin{figure}[tbp]
\begin{center}
\includegraphics[width=0.5\textwidth,clip]{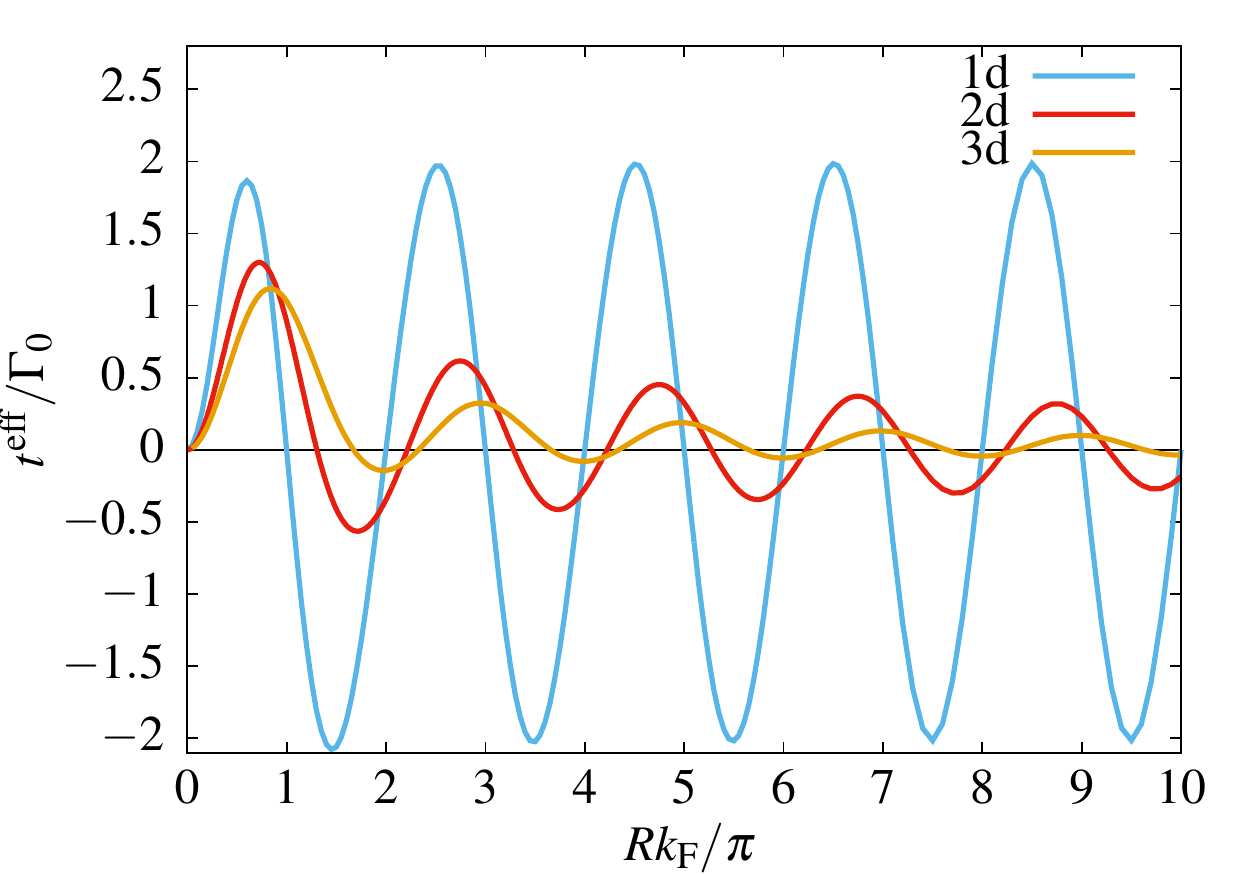}
\caption{Effective hopping element $t^\text{eff}$
for different spatial dimensions d as a function of the dimensionless distance $Rk_\text{F}/\pi$.}
\label{fig1}
\end{center}
\end{figure}

The phase shifts at the Fermi energy can be extracted from the local single- particle Green functions.
For the full problem including the counter term, 
the Green function takes the form
\begin{eqnarray}
 G_{\mu}(z,\vec{R})&=&\left(z-\e_0^f -\Delta_{\mu}(z,\vec{R})  -\Sigma_\mu^U(z)+s_\mu\frac{t^\text{eff}}{2} \right)^{-1}
 \nonumber \\
\end{eqnarray}
where $\Sigma_\mu^U(z)=\Sigma_\mu^U[G]$ denotes the correlation self-energy that is given
by a functional of the Green function \cite{LuttingerWard1960},
and
\begin{eqnarray}
\label{eqn:gammapm}
\Delta_{\mu}^\pm(z,\vec{R}) &=& V_\mu^2(\vec{R})\int_{-D}^{D} d\w \frac{\bar \rho^{(\pm)}_\mu(\w,\vec{R})}{z-\w}
\\
\label{eqn:gamma-full}
\Delta_{\mu}(z,\vec{R}) &=& \int_{-D}^{D} \frac{d\w}{\pi} \frac{\Gamma_\mu(\w,\vec{R})}{z-\w}
\nonumber \\
&= &
 \Delta_{\mu}^{(+)}(z,\vec{R}) +\Delta_{\mu}^{(-)}(z,\vec{R}) \, .
\end{eqnarray}
For $T\to 0$, the spectral function always takes the form \cite{Langreth1966,Anders1991}
\begin{eqnarray}
\rho^f_\mu(0,\vec{R}) &=& \lim_{\delta\to 0} \frac{1}{\pi} \Im  G_{\mu}(0-i\delta,\vec{R})\nonumber \\
& =& \frac{1}{\pi \Gamma_\mu(0)}
\sin^2 \delta_\mu
\end{eqnarray}
relating the scattering phase $\delta_\mu$,
\begin{eqnarray}
\label{equ-phase-shift}
\cot(\delta_\mu) &=& \frac{\e_0^f +\Re \Delta_{\mu}(0) +\Re\Sigma^U_\mu(0) -s_\mu\frac{t^\text{eff}}{2}}{  \Gamma_\mu(0)}
\end{eqnarray}
to the ratio of the real and imaginary part of the inverse Green function \cite{Langreth1966,Anders1991}. 
Note that  the Fermi-liquid property $\Im\Sigma^U(0\pm i \delta)=0$ at $T=0$ has entered as well as a coupling $|\Gamma_\mu(0)| >0$.

In general, this a complicated problem determining $t^\text{eff}$
by the condition $\delta_\mu=const$.
Therefore, we restict ourselves to a locally P-H symmetric impurity $\e_0^f+U/2=0$
that implies a Hartree term $\Re\Sigma^U(0)=U/2$.
Since $\delta_\mu=\pi/2$ independent of $\Gamma_\mu(0)$, the nominator must vanish which leads to the condition
\begin{eqnarray}
\label{eqn:t-eff-analytical}
t^\text{eff}(\vec{R})&=& 2s_\mu V_\mu^2(\vec{R})\int_{-D}^{D} d\w \frac{\bar \rho_\mu(\w,\vec{R})}{\w}
\\
&= &2 s_\mu \Re\left(\Delta_{\mu}(0,\vec{R})\right)
= 2 s_\mu \Re\left(\Delta^{(-)}_{\mu}(0,\vec{R})\right).
\nonumber 
\end{eqnarray}

\begin{figure}[t]
\begin{center}
\includegraphics[width=0.5\textwidth,clip]{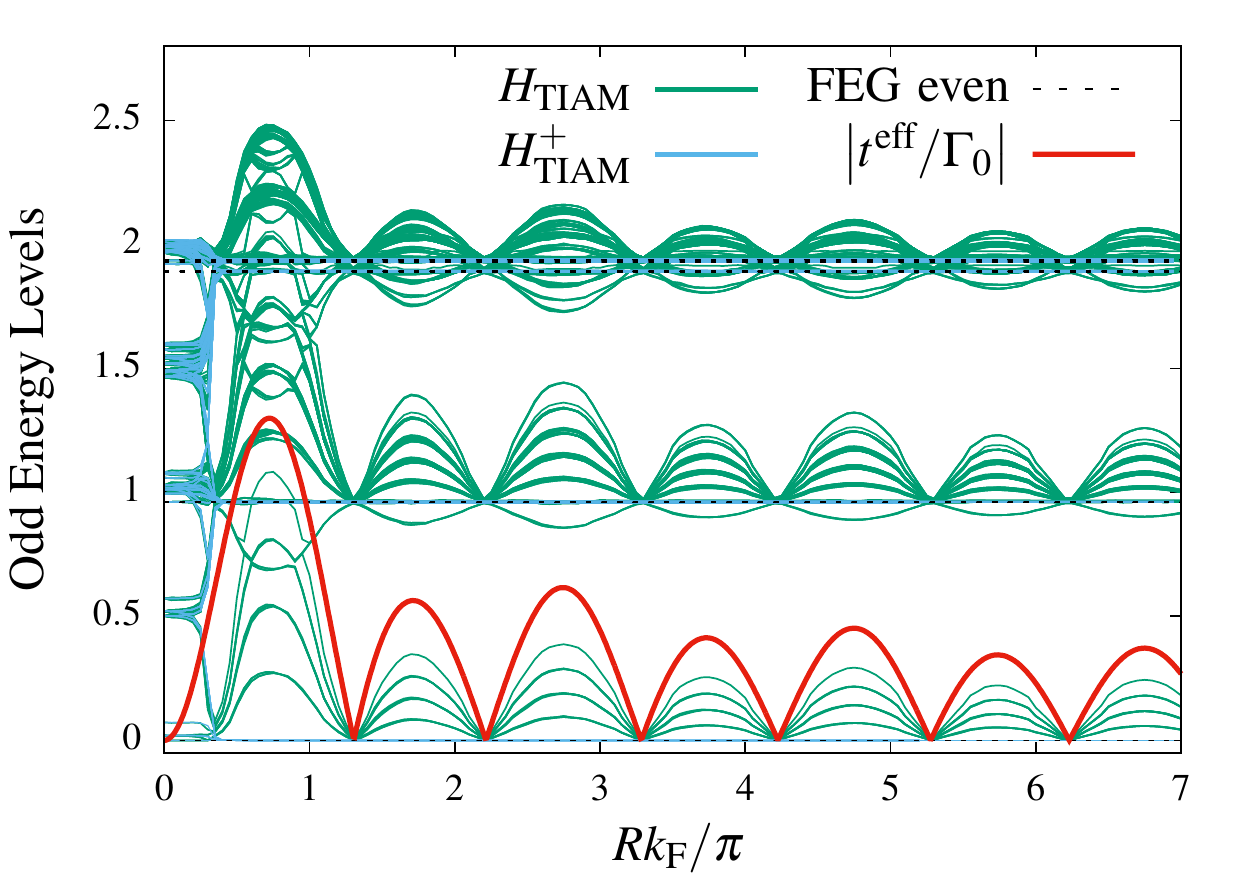}
\caption{Low temperature fixed point spectrum for an odd number of NRG-iterations 
as a function of
 the dimensionless distance $x=Rk_\text{F}/\pi$ for an isotropic 
linear dispersion $\epsilon_{\vec{k}}^c$ in two dimensions. Comparison between the full Hamiltonian (green line), the
symmetric fraction (blue line) and the free electron gas (dashed line). In the strong coupling fixed point the energy
levels 
corresponding to an odd number of iterations are comparable with the even ones from the free electron gas, since one
conduction electron degree of freedom is locked into a singlet with the impurity electron
\cite{Jones1988,Jones1989}. NRG parameters are:
Disctretization $\Lambda=1.5$, number of the kept states $\text{N}_\text{s}=4000$, $U/\Gamma_0=10$, $\epsilon^f/\Gamma_0=-5$, $t/\Gamma_0=0$, $D/\Gamma_0=10$.}
\label{fig2}
\end{center}
\end{figure}

In order to set the stage for the full NRG calculations below, we assume a constant DOS $\rho(\e)=\rho_0= 1/2D$
and an  isotropic linear dispersion $\epsilon_{\vec{k}}^c=v_\text{F}\left(|\vec{k}|-k_\text{F}\right)$,
where $v_\text{F}$ is the Fermi velocity and $k_\text{F}$ the Fermi wave-vector.
The evaluation of Eq.\ \eqref{eq:def-gamma-mu} can be performed analytically \cite{Jones1988,Jones1989,Lechtenberg2014,Borda2007}
for different spatial dimensions
\begin{align}
\label{eqn:N1D}
&1\text{d}:V^2_\mu(\vec{R})\bar \rho_{\mu}(\e,\vec{R}) =2\rho_0\left\{1+s_\mu\text{cos}\left[Rk_\text{F}\left(1+\frac{\epsilon}{D}\right)\right]\right\},\\
\label{eqn:N2D}
&2\text{d}:V^2_\mu(\vec{R}) \bar \rho_{\mu}(\e,\vec{R}) =2\rho_0\left\{1+s_\mu J_0\left[Rk_\text{F}\left(1+\frac{\epsilon}{D}\right)\right]\right\},\\
\label{eqn:N3D}
&3\text{d}:V^2_\mu(\vec{R})\bar \rho_{\mu}(\e,\vec{R}) =2\rho_0\left\{1+s_\mu\frac{\text{sin}\left[Rk_\text{F}\left(1+\frac{\epsilon}{D}\right)\right]}{Rk_\text{F}\left(1+\frac{\epsilon}{D}\right)}\right\},
\end{align}
where $R=|\vec{R}|$ is the absolute distance between the impurities and $J_0(x)$ denotes the zeroth Bessel function of the first kind.

We defined $\Gamma_0= V^2\pi \rho_0$ and plot
the  effective hopping parameter $t^\text{eff}(R)$ as function of the dimensionless distance $x=Rk_\text{F}/\pi$
for different spatial
dimensions  in Fig. \ref{fig1}.

\section{Application of the effective tunneling term}
\label{sec:results}
%----------------------------------

\subsection{Study of the low temperature fixed point}
\label{sec:fixedpoint}

\begin{figure}[b]
\begin{center}
\includegraphics[width=0.5\textwidth,clip]{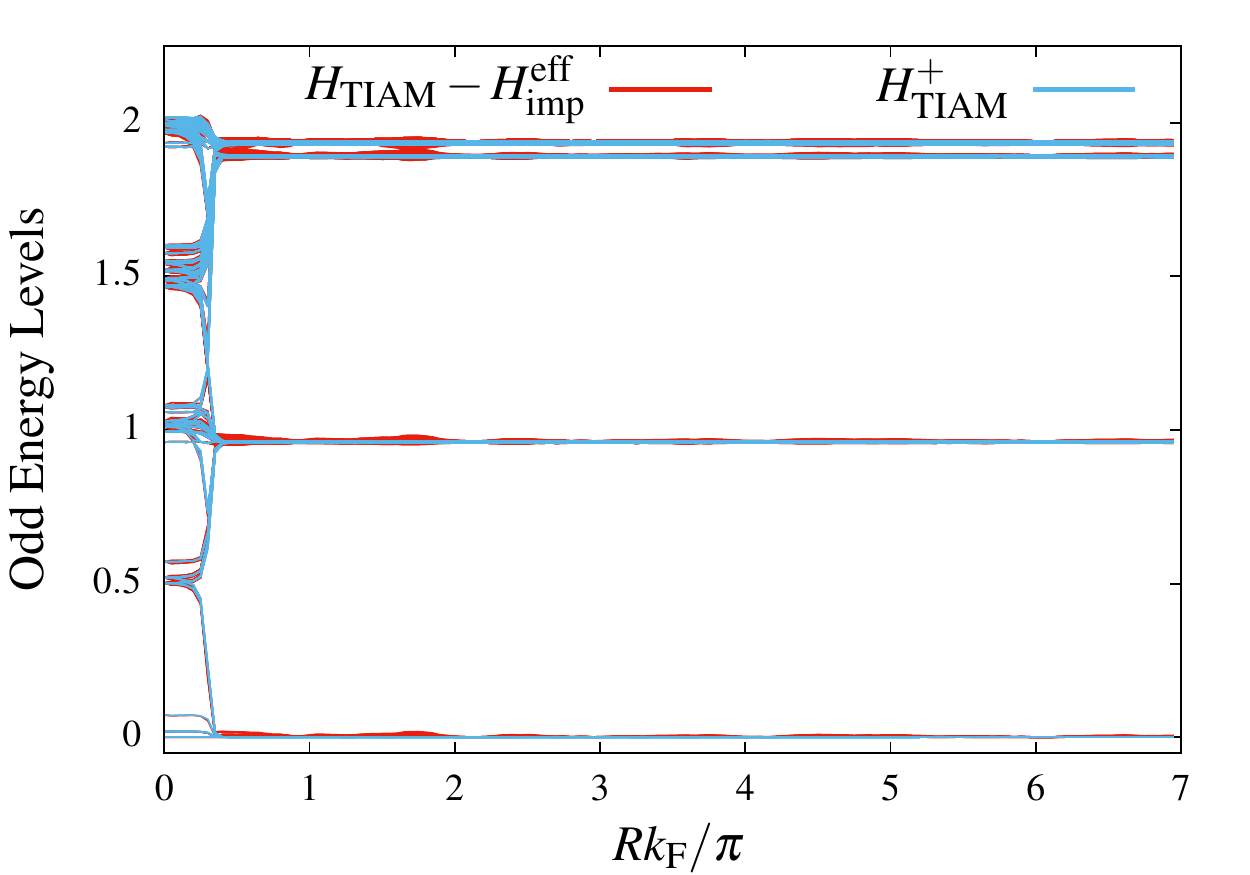}
\caption{Low temperature fixed point spectrum for an odd number of NRG-iterations in dependence of the dimensionless distance $x=Rk_\text{F}/\pi$ for an isotropic 
linear dispersion $\epsilon_{\vec{k}}^c$ in two dimensions. Comparison between the full Hamiltonian minus the effective tunneling (red line) and the
symmetric fraction (blue line). NRG parameters: as in Fig. \ref{fig2}.}
\label{fig3}
\end{center}
\end{figure}

The effective scattering terms generated by the P-H asymmetric densities of states 
$\bar\rho_\mu^-(\epsilon,\vec{R})$
influence the fixed point spectrum of the full Hamiltonian $H_\text{TIAM}$
in Eq.\;\eqref{eqn:htiam}.
For the analysis the distance dependence of these scattering terms,
we examine the fixed point properties of the full Hamiltonian $H_\text{TIAM}$ in Eq.\ \eqref{eqn:htiam}, the P-H symmetric fraction $H^+_\text{TIAM}$ in Eq.\ \eqref{eqn:Heff} and
the FEG $H_\text{host}$ in Eq.\ \eqref{eqn:hhost} by means of NRG \cite{Wilson1975,Krishna-murthy1980I,Krishna-murthy1980II,Bulla2008} 
in the P-H symmetric case $\e_f+U/2=0$.

\begin{figure}[b]
\begin{center}
\includegraphics[width=0.5\textwidth,clip]{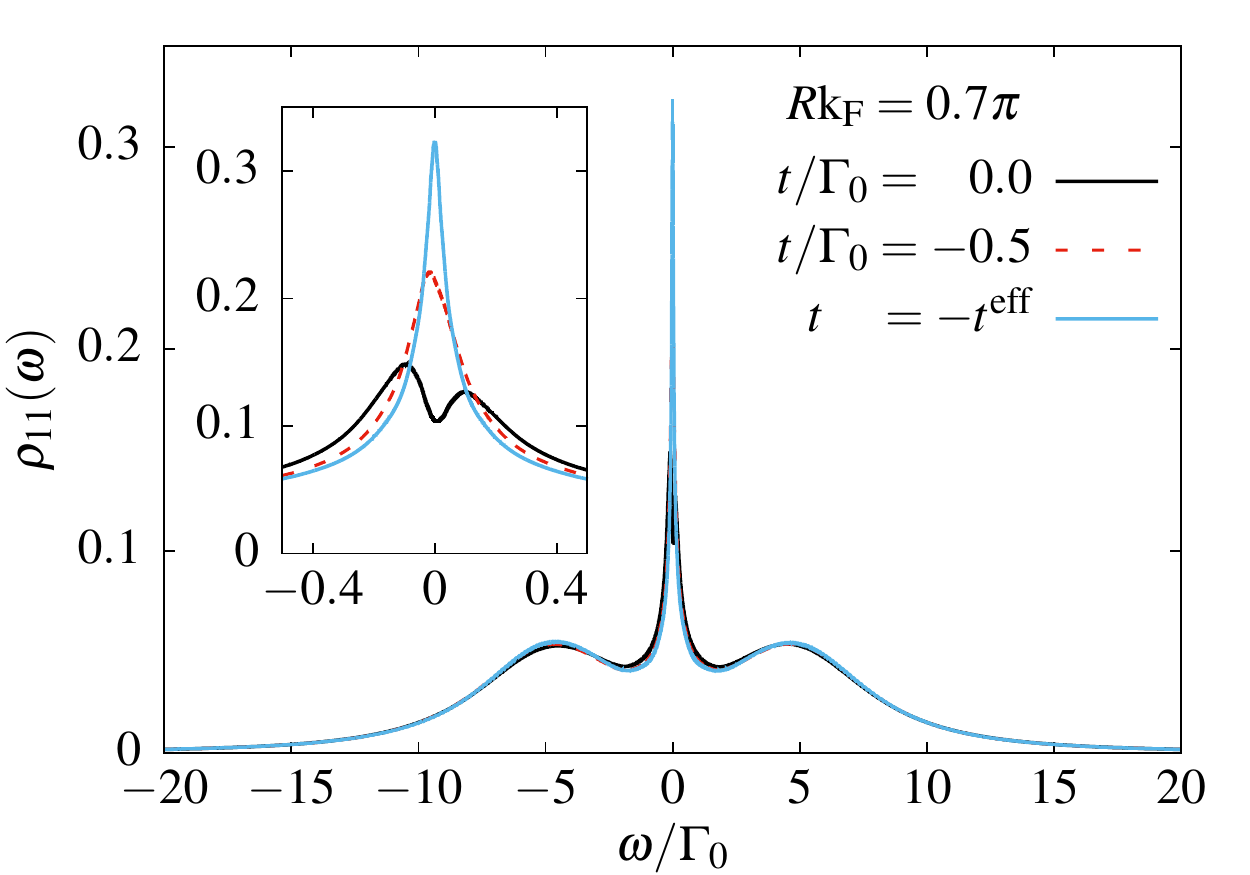}
\caption{One particle spectral function of the impurities for different tunneling parameters and a spatial separation
of $Rk_\text{F}/\pi=0.7$ on a two dimensional surface. The effective tunneling $t^\text{eff}(Rk_\text{F}/\pi=0.7)=-1.2954\Gamma_0$
leads to an P-H asymmetric gap formation around $\omega=0$ that can be effaced with an additional hopping element $t=-t^\text{eff}$.
NRG parameters: as in Fig. \ref{fig2} but with $\Lambda=2$ and $D/\Gamma_0=30$.}
\label{fig4}
\end{center}
\end{figure}

Fig.\ \ref{fig2} shows the low-temperature NRG FP spectrum 
for a two-dimensional host as function of the dimensionless distance $x=Rk_\text{F}/\pi$
and at odd iteration for the interacting Hamiltonians and even iteration of the FEG.
Since the odd conduction band decouples for $R\to 0$ \cite{Esat2016,Vojta2002,Satoshi2006} , 
the level flow of $H_\text{TIAM}$ matches those of $H_\text{TIAM}^+$ that 
is very different to the flow
of the FEG. This FP  is well understood: 
only one half of the local triplet state can be screened by the conduction electrons and the system remains in
an underscreend Kondo fixed point \cite{Esat2016,Satoshi2006,Vojta2002,Lechtenberg2017}.

In this paper, however,  we will focus on finite distances.
The low temperature FP spectrum of $H_\text{TIAM}^+$ at odd iterations coincides with those of the free electron gas at even iterations
in contrast to those of the full Hamiltonian where the influence of the effective potential scattering terms lifts the
degeneracies caused by the P-H symmetry of  $H_\text{TIAM}^+$.
The periodic structure of the fixed point spectrum of the full Hamiltonian as function of distance 
traces the oscillation of $t^\text{eff}(\vec{R})$
defined by Eq.\ \eqref{eqn:t-eff-analytical} which is
also added to Fig.\ \ref{fig2} as red solid curve.
Note that those distances where  $t^\text{eff}(\vec{R})$
vanished, the FP spectra of $H_\text{TIAM}$ matches the one for
the P-H symmetric free electron gas.

In order to check the accuracy  of the predicted  effective hopping element
we need to prove that $t^\text{eff}(\vec{R})$
is able to compensate the scattering terms due to P-H asymmetry in $\bar\rho_\mu(\epsilon,\vec{R})$
so that the FP spectra of $H_\text{TIAM}-H^\text{eff}_\text{imp}$ and $H^+_\text{TIAM}$ become identical.
These two FP spectra are depicted in Fig.\ \ref{fig3}. 
The oscillations of the energy levels disappear in $H_\text{TIAM}-H^\text{eff}_\text{imp}$
as a consequence of the counter term $H^\text{eff}_\text{imp}$ and both fixed 
point spectra coincide up to NRG discretization errors
that would require a small correction of analytically calculated $t^{\rm eff}$ in order to obtain a perfect cancelation.

The single-particle spectral function of the impurities depicted in 
Fig.~\ref{fig4} proves the restoring of the P-H symmetry around the Fermi energy
by adding the additional 
counter term. 
In the absence of the counter term, the spectral function (black line)
is asymmetric and the Kondo peak is split \cite{Satoshi2006,Sakai1990,Schmitt2012} as can be seen in the inset of Fig.~\ref{fig4}.
By compensating the intrinsic, effective tunneling, the splitting of the Kondo resonance 
vanishes (light blue line).

\subsection{Resorting the Varma and Jones quantum critical point}
\label{sec:QCP}

\subsubsection{Local P-H symmetry on the impurities}
\label{sec:PH-symm-VJ}

The Varma and Jones (VJ)  QCP is inevitably stable in the presence of the P-H symmetry of the first type, as becomes apparent 
by describing the Fermi-liquid phase in terms of the phase 
shifts in the even and odd channels at zero energy \cite{Affleck1995}. 
Making use of the
symmetry transformation \eqref{eqn:phs1} in combination with the boundary conditions for incoming and outgoing conduction electrons
\begin{align}
 c^{(\dagger)\,\text{out}}_{e/o}(\epsilon)=e^{(-)2i\delta_{e/o}} c^{(\dagger)\,\text{in}}_{e/o}(-\epsilon),
\end{align}
pins the possible phase shift to $\delta_{e/o}=0\lor \pi/2$. As a result there is a QCP separating the 
Kondo-screening phase ($\delta_{e/o}=\pi/2$) and the inter-impurity singlet phase ($\delta_{e/o}=0$), whereas absence 
of the P-H symmetry of the first type allows a general phase shift $\delta_{e/o}\in[0,\pi/2]$ with a smooth crossover from $0$ to $\pi/2$.

In the preceding section \ref{sec:fixedpoint} we established
the restoration of the P-H symmetry of the first kind  in the FP spectrum by a counter  $H^\text{eff}_\text{imp}$.
For a vanishing $t^{\rm eff}$, the FP of $H_\text{TIAM}^{\rm eff}$  turns out to be already P-H symmetric.

\begin{figure}[t]
\begin{center}
\includegraphics[width=0.5\textwidth,clip]{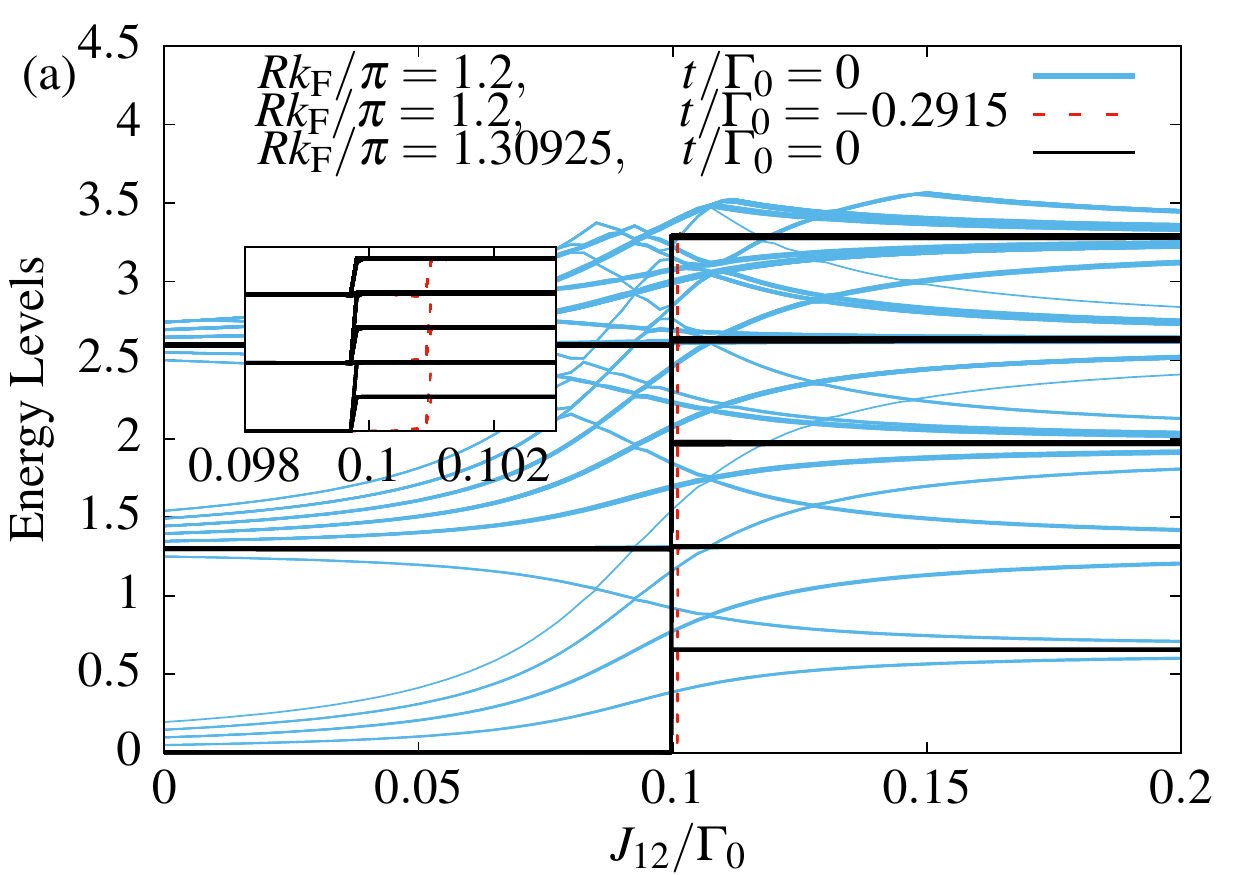}
\includegraphics[width=0.5\textwidth,clip]{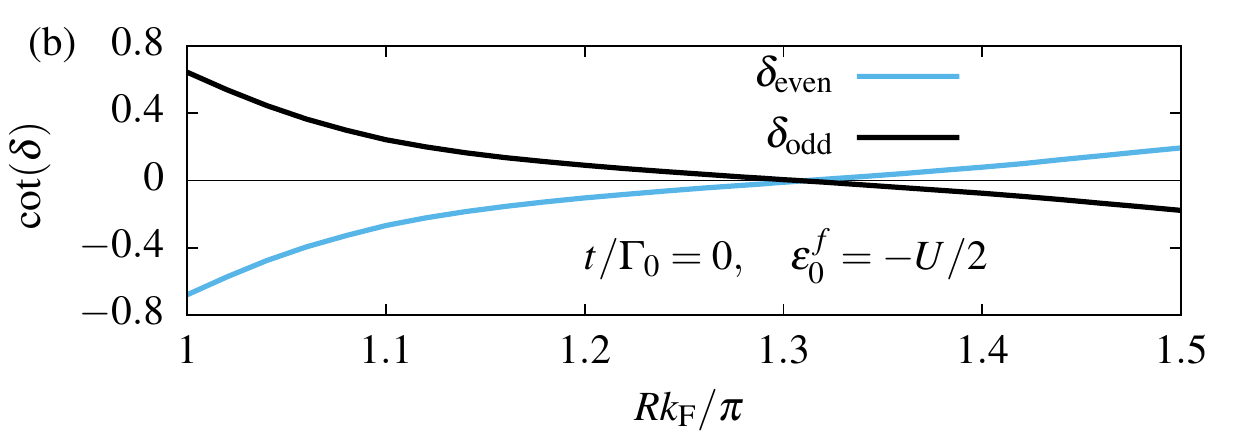}
\caption{(a) Development of the low temperature fixed point spectrum with increasing antiferromagnetic
inter-impurity spin exchange $J_{12}/\Gamma_0$ in two dimensions. A smooth crossover appears for a general P-H asymmetric
Hamiltonian with $t=0$ and $t^\text{eff}(R)\not=0$ (blue lines) by contrast with a quantum phase transition for the special case $t^*+t^\text{eff}(R)=0$ (red lines)
as well as $t=0$ and $t^\text{eff}(R^*)=0$ (black lines). The inset depicts a zoom around the critical value $J_{12}^c/\Gamma_0$.
(b) Scattering phase in the even and odd chanels for a P-H symmetric impurity, plottet against the impurity distance. The QCP 
exists for $\delta_e=\delta_o=\pi/2$ at $R\approx1.30925$.
NRG parameters:  as
in Fig. \ref{fig2} but with $\Lambda=2$.}
\label{fig5}
\end{center}
\end{figure}

In order to prove the presence of the QCP, we 
added a direct Heisenberg exchange interaction $J_{12}\vec{S}_1\vec{S}_2$ to the full
two impurity Hamiltonian,
\begin{eqnarray}
H'_\text{TIAM}(J_{12})&=& H_\text{TIAM} + J_{12}\vec{S}_1\vec{S}_2.
\end{eqnarray}
Fig.~\ref{fig5}(a) depicts three different FP level spectra 
as function of $J_{12}$:  for $Rk_F/\pi=1.2$  with (red dashed line) 
and without (light blue solid line) a counter term
and at the special  distance $R k_F/\pi=1.30925$ (black solid line)
In accordance with the literature, the transition from the Kondo regime
($J_{12}\rightarrow -\infty$) to the inter-impurity singlet regime ($J_{12}\rightarrow \infty$) 
is continuous for a generic distance 
such as $Rk_F/\pi=1.2$ (blue lines in Fig.\ \ref{fig5}(a))
without an additional counter term.

As demonstrated by the FP spectra, the Varma-Jones QCP can be restored by adding a direct tunneling 
$t^*=-t^\text{eff}(R)$. 
The level flow jumps discontinuously from one to another FP spectrum at a critical coupling $J^c_{12}$
revealing clearly the QCP.
Evaluating Eq.\ \eqref{eqn:t-eff-analytical} for this distance yields $t^*(Rk_F/\pi=1.2)/\Gamma_0=-0.2915$.

Alternatively, the distance can be varied to values $R^*$ such that $t^\text{eff}(R^*)$
vanished and hence $\cot \delta_e=\cot \delta_o=0$.
Fig.\  \ref{fig5}(b) shows the distance dependency of the scattering phase
using the model parameters of Fig.\ \ref{fig2}. We determined the
shortest finite distance for which this condition is fulfilled as $R^*k_\text{F}/\pi\approx 1.30925$.
For this distance $R^*$, we scan the  FP level flow as function of $J_{12}$ and add the
results to Fig.~\ref{fig5}(a) as solid black line. Clearly, we also find a QCP at almost the same
critical value for  $J_{12}$. The inset in  Fig.~\ref{fig5}(a) resolves the very small distance
dependent shift of the critical value compared to the case of
the generic distance $Rk_F/\pi=1.2$ with the additional counter term.

\subsubsection{Local P-H asymmetry on the impurities}

Now we proceed to the generic case 
where also the local P-H symmetry on the impurities is broken
but the parity remains conserved.
For a fixed $U$, the single particle energy is given by the onsite energy $\epsilon^f=-U/2+\Delta\epsilon$
where $\Delta\epsilon$ parameterized 
its deviation from the P-H symmetric point. Leaving $\epsilon^f_0=-U/2$, the addition
term 
\begin{align}
 H_{\Delta \epsilon}=\Delta\epsilon\sum_{\sigma}\left(f^\dagger_{e,\sigma}f_{e,\sigma}+f^\dagger_{o,\sigma}f_{o,\sigma}\right)
\end{align} 
accounts for the local  P-H asymmetry on the impurities. It 
leads to  potential scattering parameter in the form of $K_e\not=-K_o$.
Since the absolute value of the scattering terms in the even and in the odd channel does not coincide, it is not possible to cancel both terms simultaneously by 
introducing  a direct
tunneling term or varying the spatial separation.

\begin{figure}[t]
\begin{center}
\includegraphics[width=0.5\textwidth,clip]{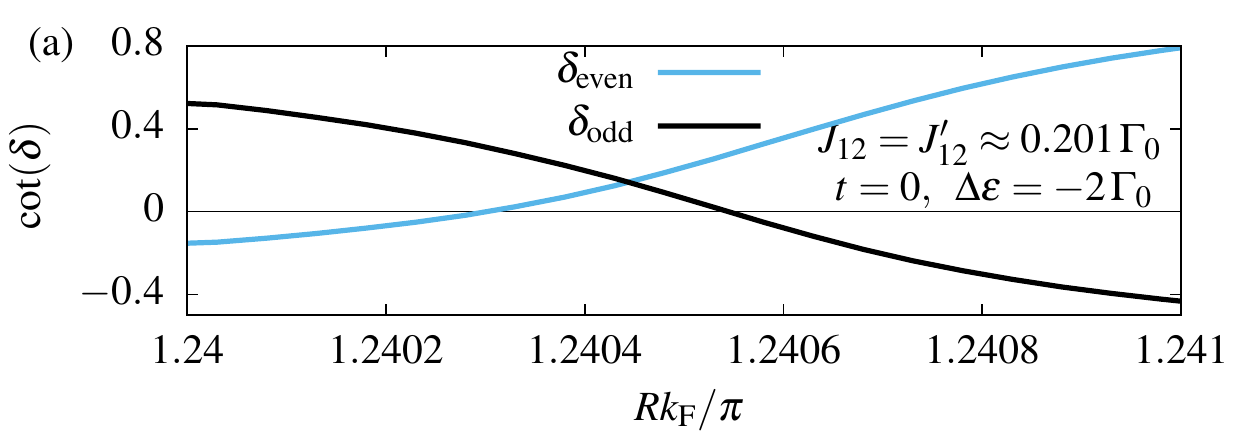}
\includegraphics[width=0.5\textwidth,clip]{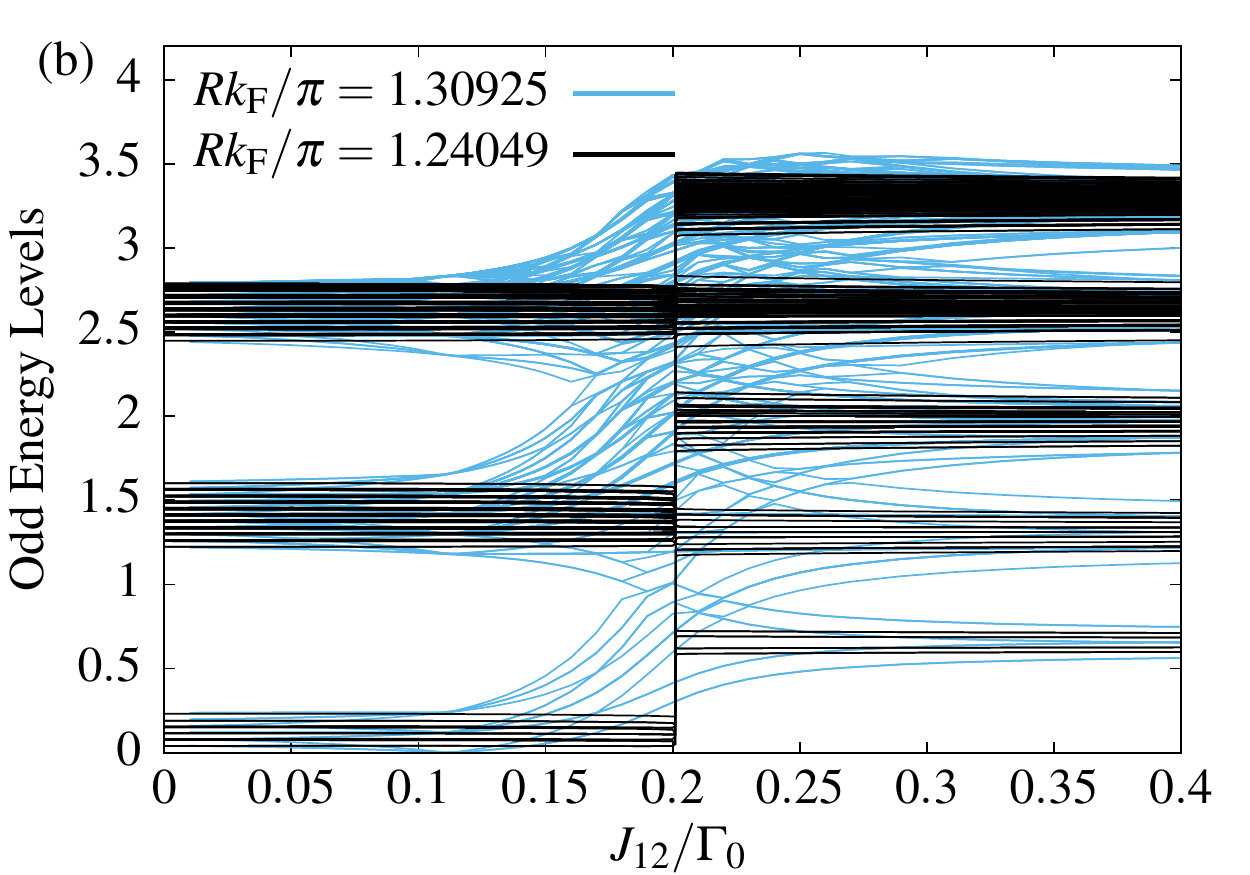}
\caption{
(a): Scattering phase in the even and odd channel for a P-H asymmetric impurity
as function of $R$.  (b): Low temperature FP spectrum as function of
$J_{12}/\Gamma_0$ and $\Delta\epsilon/\Gamma_0=-2$
for   $R^*(\Delta\epsilon=0)k_\text{F}=1.30925\pi$ (blue lines) and 
 $R^*(\Delta\epsilon=-2)k_\text{F}=1.24049\pi$ (black lines).
NRG parameters: as in Fig. \ref{fig5}.
 }
\label{fig6}
\end{center}
\end{figure}

We will demonstrate that the VJ QCP can be restored 
by changing the low energy scattering terms such that they 
generate identical scattering phases in the even and the odd channel,
i.\ e.\ $\delta_e=\delta_o$.
Zhu and Varma \cite{Zhu2006} pointed out that the scattering phase  acquires  an additional
contribution $\Delta \delta_\mu = −\tan^{-1}(\pi \rho_0 K_\mu)$  in the SC FP  caused by a P-H asymmetry.

Since neither $\Delta \delta_\mu$ nor $K_\mu$ is directly accessible in the NRG, we use a different strategy 
that is directly 
based on the NRG FP spectra. 
Close to the P-H symmetric point, the difference between the 
lowest  single-particle excitation relative to the NRG ground state, $E^1_{\mu}$, 
with an even parity ($\mu=e$)  and an odd parity ($\mu=o$),  
\begin{eqnarray}
\Delta\omega_0 &=& E^1_e - E^1_o,
\end{eqnarray}
is proportional to the difference of the 
phase shifts.

Tuning the  inter-impurity spin exchange $J_{12}$
generically drives the system continuously from a SC to a local singlet FP
and $\Delta\omega_0$ changes continuously. For a sharp transition,
$\Delta\omega_0$ must vanish at  the critical coupling $J_{12}^c$ 
\begin{eqnarray}
J_{12}^\prime&=&\lim\limits_{\delta \rightarrow 0}(J_{12}^c+\delta).
\end{eqnarray}
Note that the phase shifts at $J_{12}=J_{12}^c$ are not defined. 
Since the critical value $J_{12}^c$ is unknown apriori, it
leads to the self-consistency condition:
\begin{align}
\label{eqn:DeltaW}
 \Delta\omega_0\left(\Delta\epsilon,R^*,t^*,U,J_{12}^\prime\right)=0.
\end{align}
This equation is solved iteratively.

As starting point,
we choose the critical value
$J_{12}^c$ for the local P-H symmetric case, i.\ e.\ $\Delta\e=0$. 
Then we compute $\Delta\omega_0$ as function of $R$ ($t$ respectively) and determine
the roots for $R^*_1$ ($t^*_1$) for constant $t$ ($R$ respectively).
In the next step, we determine the $J_{12}^{\prime2}$ at the midpoint of the crossover regime.
 Inserting $J_{12}^{\prime2}$ into Eq.\;\eqref{eqn:DeltaW} results in new
$R^*_2 (t^*_2)$. This steps are iterated until convergence is achieved.

Starting at the critical distance $R^*_1=1.30925\pi/k_F$,
obtained for $\Delta\epsilon=0, U/\Gamma_0=10,\quad t/\Gamma_0=0$
in Sec.\ \ref{sec:PH-symm-VJ}, this procedure converged after four iterations to 
$R^* k_\text{F}=1.24049\pi$ to a precision of
5 digits.

Fig.\ \ref{fig6}(a) displays the even and odd scattering phases in the last 
iteration,  i.\ e.\ for the critical spin exchange $J^c_{12}$, as function of the distance.
This convincingly demonstrates  the consistency of our approach:
Fixing the last value of $J^c_{12}$, the point of coincidence  of the two scattering phases
agrees perfectly with the critical $R^*(\Delta\epsilon/\Gamma_0=-2)k_\text{F}=1.24049\pi$
obtained by the iteration procedure.

In order to prove that the VJ QCP is
really restored for this set of parameters, we present the 
FP level flow as function of the  coupling $J_{12}$ in 
Fig.\ \ref{fig6}(b) for the starting distance 
starting distance $R^*_1k_\text{F}=1.30925\pi$ (blue lines)
and the final distance  $R^*$ (black lines). While only a crossover is observed
for $R^*_1$, clearly the VP QCP is restored at the final distance $R^*$ even for
$\Delta\epsilon/\Gamma_0=-2$. The additional
term $t^{\rm eff}$ is not needed. Note the FP level flow in both phases: the different
magnitude of the P-H symmetry breaking scattering term in both phases
is clearly visible.

\subsection{Splitting of the RKKY interaction in two contributions}
\label{sec:RKKY}

The RKKY interaction between two local moments with a distance $R$ apart 
is mediated by the metallic host. This effective coupling constant $J_{\rm RKKY}$ is distance 
dependent and shows the characteristic alternating signs with $2k_F$ oscillations
-- at least for 
a simplified dispersion of the conduction electrons.

Consequently, we can divide the RKKY interaction into two contributions with  opposite signs.
Extending the argument for a constant DOS  \cite{Jones1987,Lechtenberg2014}
one can show that a  P-H symmetric effective DOS  $\bar\rho^{(+)}(\e)$
can only generate a ferromagnetic RKKY interaction  $J_{\rm RKKY}^{\rm FM}$ at arbitrary distances.
Hence, the antiferromagnetic contribution
results from the breaking of the P-H symmetry of the first type
that can be parameterized by a local $t^\text{eff}$. 

Decoupling of the impurities from the effective conduction electrons allows for
an exact solution of this effective two impurity problem. 
For $t^\text{eff}=0$, the local triplet state involving both even and odd orbital
is degenerate with the singlet state given by the linear combination of both electrons in the even or both electrons in the odd state \cite{Esat2016}.
A finite $t^\text{eff}$ induces an imbalance between the mixing of these singlet states and an energy gain of $J_{\rm ex}= |t^\text{eff}|^2/U>0$
that can be interpreted as effective interaction between the two local spins 
in the local moment regime. Clearly, this local exchange mechanism always generates an antiferromagnetic interaction.

For the local P-H symmetric case, the analytic solution \eqref{eqn:t-eff-analytical}
predicts 
$t^\text{eff} \propto \rho_0 V^2$, and the Schrieffer-Wolff transformation  \cite{Schrieffer1966} 
generates
a local Kondo coupling $J_K \propto V^2/U$. Therefore, the local  exchange term can
be related to $J_K$ via
\begin{eqnarray}
J_{\rm ex} &=&  \frac{|t^\text{eff}|^2}{U} \propto U (\rho(0)J_K)^2
\end{eqnarray}
This is a generalization of the $R\to 0$ analysis of FM RKKY in an
multi-impurity model \cite{MultiImpurityAM2006} to AF contributions
for arbitrarily distances $R$.
The estimated order of magnitude of $J_{\rm ex} \propto 1/U$ agrees perfectly with
the cumbersome evaluation of a Rayleigh-Schrödinger perturbation
theory in forth order \cite{MultiImpurityAM2006}.
Our analyzes provides a much simpler understanding of the 
difference of the RKKY interaction in the two-impurity Anderson model
and in the two-impurity Kondo model.

Combining these two terms yields the total RKKY coupling $J_{\rm RKKY} = J_{\rm RKKY}^{\rm FM} + J_{\rm ex}$. 
This leads to the interesting fact that by
adding an additional inter-impurity orbital
hopping term $t$, it is possible to change the sign of the
total coupling $J_{\rm RKKY}$ in arbitrary direction. 
Typically, a tunneling term only generates a AF exchange interaction, however, 
adding a $t$ with opposite sign compared to $t^\text{eff}$ reduces the total tunneling $\bar{t}^\text{eff}=t^\text{eff}+t$
and may eventually cause a sign change to a FM $J_\text{RKKY}$.
On the other side,
starting from $t^\text{eff}=0$, i.\ e.\ a
purely FM $J_{\rm RKKY}$ and increasing $t$, induces a AF coupling
that become arbitrarily large and eventually will lead to
a sign change.

\begin{figure}[t]
\begin{center}
\includegraphics[width=0.5\textwidth]{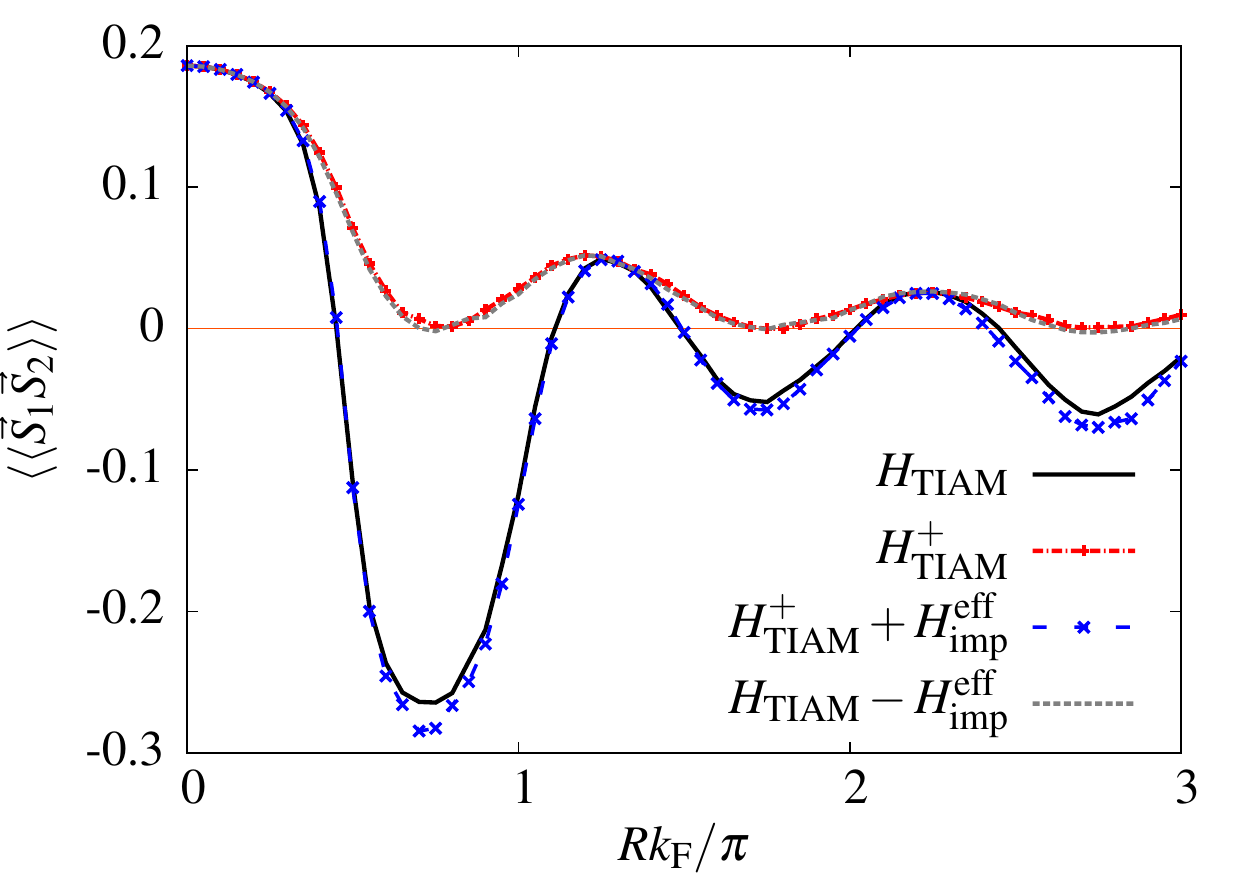}

\caption{Impurity spin-spin correlation function as function of the distance for the full TIAM Hamiltonian,
for the effective model and the symmetric part $H_{\rm TIAM}^{+}$. A featureless symmetric conduction
band with a 2d linear dispersion has been used for the locally P-H symmetric regime for $T \to 0$. 
Parameters: $U/\Gamma_0=10$, $D/\Gamma_0=100$, $\text{N}_\text{s}=4000$, $\Lambda=2$.
}
\label{fig7}
\end{center}
\end{figure}

To illustrate that the full energy dependent TIAM can be mapped 
to an effective model at low energies comprising an P-H symmetric conduction band,
generating the FM RKKY interaction, as well as
a local hopping term, which induces the AF part $J^\text{AF}_\text{RKKY}$, the impurity spin-spin correlation
function of  both models, calculated by means of NRG, is shown in Fig.\ \ref{fig7}.
The correlation function $\ll \vec{S}_1 \vec{S}_2\gg$ for  $H_{\rm TIAM}^{+}$ is purly
positive  demonstrating
that the RKKY interaction $J^\text{FM}_\text{RKKY}$ for a P-H symmetric DOS can only be FM \cite{Jones1987}.
The correlation function
of the effective model $H^+_\text{TIAM}+H^\text{eff}_\text{imp}$ agrees excellently with those of the full model 
in the short distance regime. We discuss the corrections, which occur for larger distances
due to a finite bandwidth, in the next section.

Note that there are infinitely many distances $R^*_n$ at which $t^\text{eff}=0$, so that $J_\text{RKKY} = J^\text{FM}_\text{RKKY}$ holds. At these distances, the spin-spin correlation function of all models coincided,
and the full energy dependent model with an additional direct spin-spin interaction $J_{12}$ exhibits the VJ QCP.

Note that the effective tunneling $t^\text{eff}$ which restores the P-H symmetric FP $H^+_\text{TIAM}=H_\text{TIAM}-H(t^\text{eff})$,
and the one that restores the FP of the full Hamiltonian
out of the P-H symmetric fraction $H_\text{TIAM}=H^+_\text{TIAM}+H(t^\text{eff})$, in general are not fully identical.
While for a
P-H symmetric FP, only the value at zero-frequency is relevant, and, consequently, Eq. \eqref{eqn:t-eff-analytical} is exact, corrections stemming from the derivative $d\bar \rho_\mu(\e)/d\e$
need to be taken into account to recover the FP of the full Hamiltonian.

\subsubsection{Finite bandwidth corrections}

Focusing on a  1d  conduction band with a linear dispersion for a moment,
we noticed that the amplitude of the correlation function of the effective model $H^+_\text{TIAM}+H^\text{eff}_\text{imp}$
will not decay for $t^\text{eff}$ given by Eq.\ \eqref{eqn:t-eff-analytical}.
At the distances $R_nk_\text{F}=(2n+1)\pi/2$, the Hamiltonian is P-H symmetric of the second type:
The symmetric fraction of the effective DOS $\bar{\rho}^+_\mu(R_n,\e)$ is constant and  distance independent. Furthermore, the effective tunneling is given by the
analytical expression
\begin{align}
 t^\text{eff}_{1d}(R_n)\propto\int_{-1}^1\frac{\sin(R_nk_\text{F}x)}{x}dx=2\text{Si}(R_nk_\text{F})\,,
\end{align}
where $\text{Si}(R_nk_\text{F})$ is the sine integral, which is constant for large distances
$\text{Si}(\infty)=\pi/2$.
Apparently, the effective model cannot capture the decay 
of the impurity spin-spin correlation function for large distances
and corrections to the effective model need to be taken into account.

To estimate the magnitude 
of the corrections, we analyze the resonant level model ($U=0$), where we can derive an analytic
expression for the correlation function. 
One can show that the correlation 
function is proportional to the difference
of the distance dependent occupation of the even impurity
orbital $n_e(\vec{R})$ and the odd impurity orbital $n_o(\vec{R})$,
\begin{align}
 \langle\langle\vec{S}_1\vec{S}_2\rangle\rangle^{U=0}=-\frac{3}{8}\left[n_o(\vec{R})-n_e(\vec{R})\right]^2\,.
\label{eqn:spin-corr_RML}
\end{align}
At zero temperature these occupation numbers are given by the integral of the 
analytically obtained spectral functions
\begin{align}
 n_{\mu}(\vec{R})=\int_{-\infty}^0 \frac{d\omega}{\pi} \frac{\Gamma_\mu(\omega,\vec{R})}{\left(\omega-\Re(\Delta_\mu(\omega,\vec{R}))\right)^2+\Gamma^2_\mu(\omega,\vec{R})} \,,
\end{align}
where the real and imaginary part of the hybridization function can be decomposed into the contributions
from both symmetry types: $\Gamma_\mu(\w)=\Gamma_\mu^+(\w)+\Gamma_\mu^-(\w)$ and $\Delta_\mu(\w)=\Delta_\mu^+(\w)+\Delta_\mu^-(\w)$.

In order to derive corrections, we turn to the wide band limit.
We can always find the lowest $D$ such that  
$\tilde{\e}_{\vec{k}}=\e_{\vec{k}}/D \in[-1,1]$ defines
a dimensionless band structure. From Eq.\ \eqref{eq:def-gamma-mu}, it is clear
that the energy dependence of $\Gamma_\mu(\w)$ and  $\Re(\Delta_\mu(\w))$ 
can be expressed through the dimensionless
functions $f_\mu(\w/D)$ and $F_\mu(\w/D)$: $\Gamma_\mu(\w)=\Gamma_0f_\mu(\w/D)$, 
$\Re(\Delta_\mu(\w))=\Gamma_0F_\mu(\w/D)$ and the occupation 
number can be written as
\begin{eqnarray}
n_{\mu}(\vec{R})&=
&\int_{-\infty}^0\frac{d\w}{\pi\Gamma_0} 
\frac{f_\mu(\frac{\omega}{D},\vec{R})}
{\left(\frac{\omega}{\Gamma_0}- F_\mu(\frac{\omega}{D},\vec{R}) \right)^2+ \left(
f_\mu(\frac{\omega}{D},\vec{R})\right)^2}   \, .
\nonumber \\
\end{eqnarray}
For fixed hybridization strength $\Gamma_0$, and $\Gamma_0/D\to 0$, the total spectral
weight is located around
\begin{eqnarray}
\omega_{0,\mu} &\approx& \Gamma_0F_\mu(0)+\mathcal{O}\left(\frac{\Gamma_0}{D}\right),
\end{eqnarray}
where we can neglect the correction in the wide band limit $D\to \infty$.

In the effective Hamiltonian, we include
the contributions  $\Gamma_\mu^+(\w)$ and $\Re\Delta_\mu^+(\w)$ 
exact, but $\Gamma_\mu^-(\w)$ and $\Re\Delta_\mu^-(\w)$ only up to zero-order. 
In a Taylor series, the leading corrections  are generated by the derivatives of these functions. Since
\begin{eqnarray}
\frac{d}{d\w}\Re\left(\Delta^-_\mu\left(z\right)\right)\bigr|_{\w=0}&\propto& P\int_{-1}^{1}\frac{\Gamma^-(x)}{x^2}dx=0\,,
\end{eqnarray}
where $x=\w/D$,
the leading corrections are proportional to  $\frac{d}{d\w}\Gamma^-_\mu(\w)|_{\w=0}$, at least for small 
coupling strengths $U/\Gamma_0$.
The distance dependence  enters in $\frac{d}{d\w}\Gamma^-_\mu(\w)|_{\w=0}$
differently for different spatial dimensions, but  is always proportional to $\Gamma_0/D$.
For a linear dispersion we obtain analytically 
\begin{align}
&1\text{d}:\quad\frac{d}{d\w}\Gamma^-_\mu\left(\w\right)\bigl|_{\w=0}\propto\frac{Rk_\text{F}\Gamma_0}{ D}\,,\\
&2\text{d}:\quad\frac{d}{d\w}\Gamma^-_\mu\left(\w\right)\bigl|_{\w=0}\propto\frac{\sqrt{Rk_\text{F}}\Gamma_0}{ D}\,,\\
&3\text{d}:\quad\frac{d}{d\w}\Gamma^-_\mu\left(\w\right)\bigl|_{\w=0}\propto\frac{\Gamma_0}{ D}\,.
\label{corrections-3d}
\end{align}
In the limit case of an infinite bandwidth $\Gamma_0/D\rightarrow0$, the effective tunneling determines the AFM part of the RKKY interaction on all length scales in any dimension.

\begin{figure}[t]
\begin{center}
\includegraphics[width=0.5\textwidth]{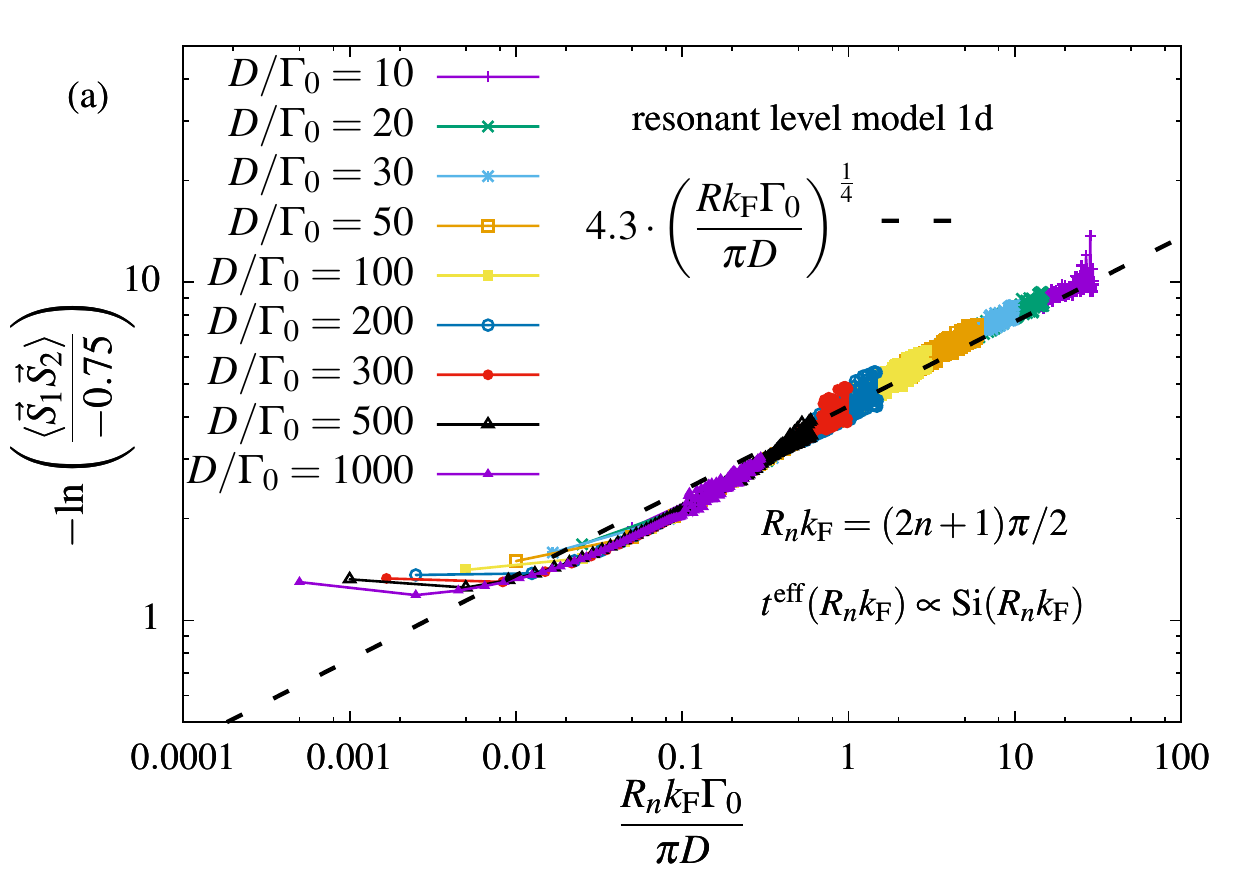}
\includegraphics[width=0.5\textwidth]{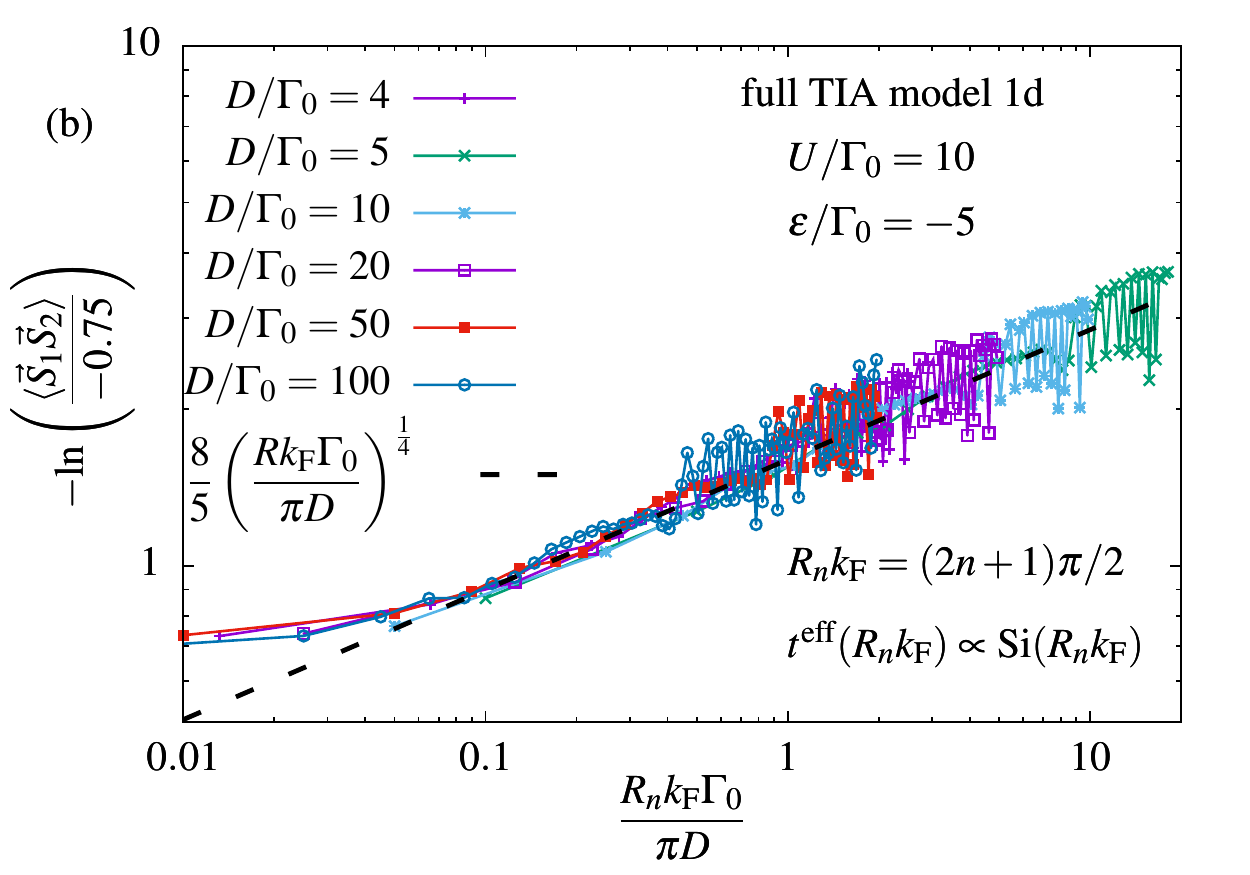}
\caption{Impurity spin-spin correlation as function of $x=(R_nk_\text{F}\Gamma_0)/(\pi D)$ 
with $R_nk_\text{F}=(2n+1)\pi/2$.
(a) Correlation function for  the resonant level model 
($U=0$) calculated using Eq.\ \eqref{eqn:spin-corr_RML} for a 1d linear dispersion.
(b) Correlation function 
for TIAM with $U/\Gamma_0=10,\e/\Gamma_0=-5$
calculated using  the NRG with $\text{N}_\text{s}=4000$ and $\Lambda=3$.
}
\label{fig8}
\end{center}
\end{figure}

These theoretical considerations are backed by a comparison of 
analytical calculations for the two-impurity resonant level model ($U=0$) in Fig.\ \ref{fig8} (a) and 
a full NRG study of the spin-spin correlation function for a finite $U/\Gamma_0=10\,$ in Fig.\ \ref{fig8} (b)
in 1d.
Fig.\ \ref{fig8} shows the correlation as function 
of $x=(R_nk_\text{F}\Gamma_0)/(\pi D) \propto 
d\Gamma^-_{1\text{d},\mu}/d\w(\w=0)$.
In order to extract the power-law of the universal corrections, we
logarithmically plot the antiferromagnetic correlation function normalized to its maximum value of $-0.75$.
Panel (a) depicts the evaluation of Eq.~\eqref{eqn:spin-corr_RML} for the resonant level model, 
whereas panel (b) shows the results for the TIAM at finite $U/\Gamma_0$, calculated via the NRG. The figure combines the 
scans for many different values of the band width at the discrete distance $R_nk_\text{F}=(2n+1)\pi/2$.
Although  the effective tunneling is nearly constant, the universality with respect to the scaling
variable $x$ is clearly demonstrated. For $x\to 0$, the correlation function approaches a finite value
above its theoretical minimum. While the correlation function is constant for small $x$
the corrections become clearly visible for $0.1<x$. 
Phenomenological, we found that a  powerlaw fit 
$\propto x^{1/4}$ agrees remarkably with the data.
Since the effective tunneling is nearly constant, only the corrections lead to a decay,
wherefore the correlation is a universal function of the parameter that characterizes the strength of these corrections.

\subsubsection{$U$-dependency of the RKKY interaction}

\begin{figure}[t]
\begin{center}
\includegraphics[width=0.5\textwidth]{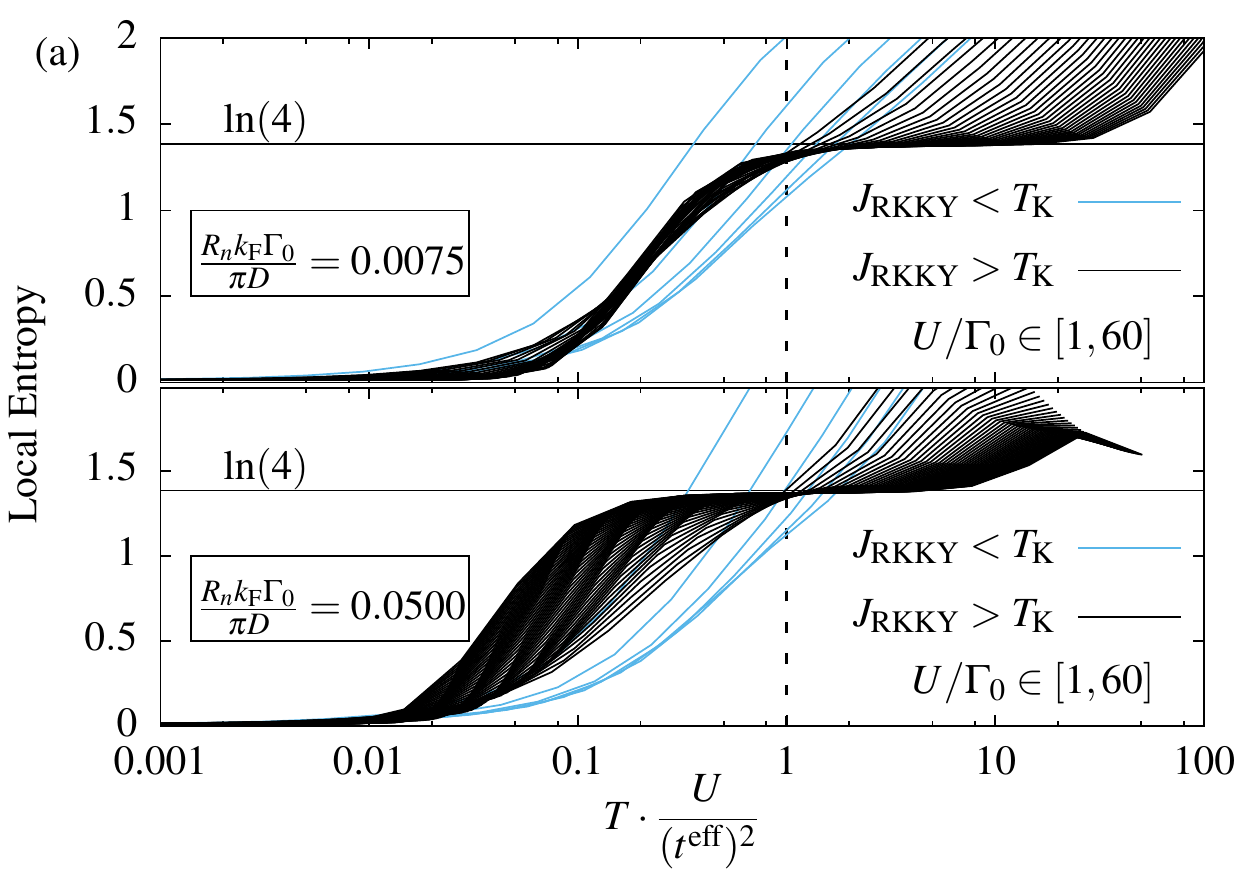}
\includegraphics[width=0.5\textwidth]{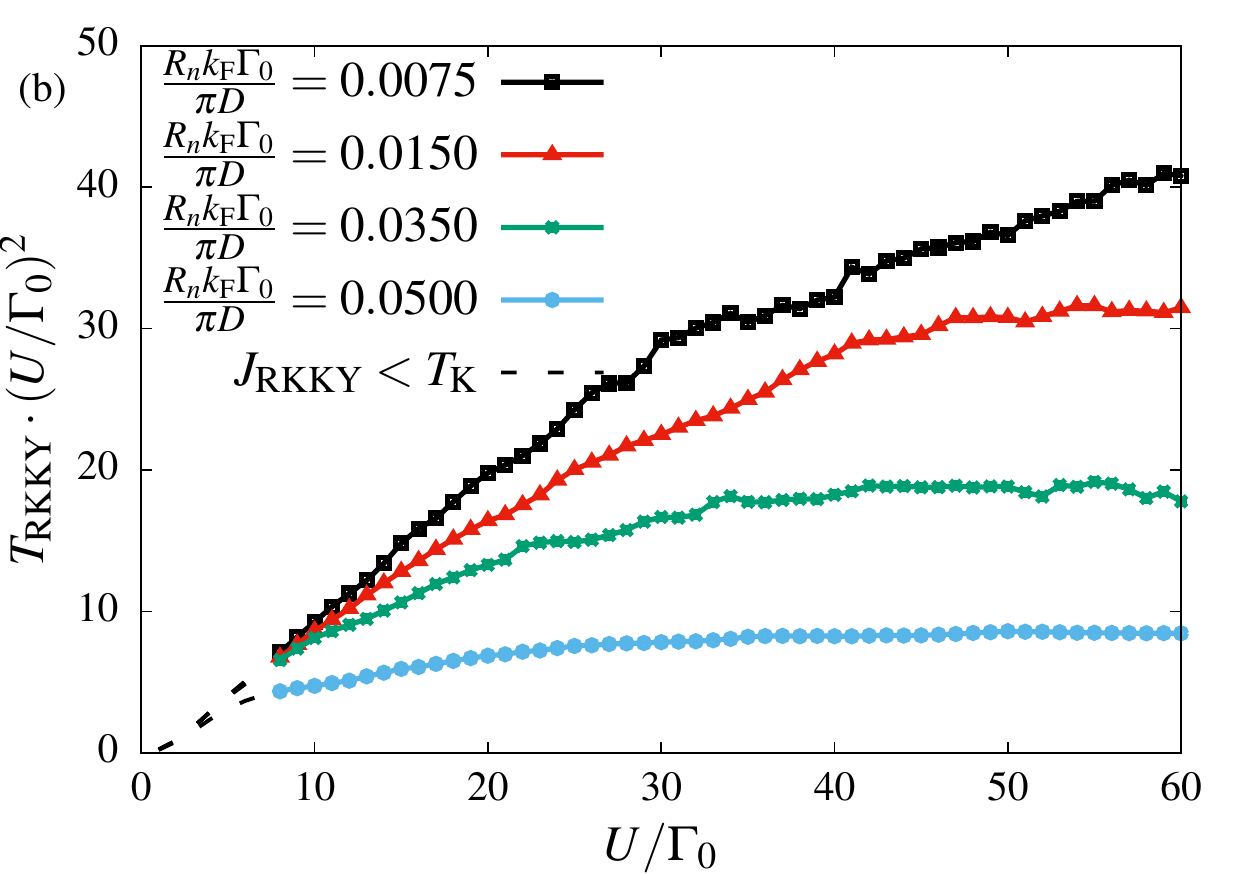}

\caption{(a) Local entropy of the impurities as function of the dimensionless temperature $t=T U/t^2_\text{eff}$
for two different effective distances $R_n k_\text{F}\Gamma_0/D$. 
The different lines represent different coupling strengths $U/\Gamma_0$ in a range of $1<U/\Gamma_0<60$.
The crossover temperature from the four-fold degenerate LM regime
a singlet with entropy $\ln(1)$ determines 
the energy scale $T_\text{RKKY}$ of the RKKY interaction.
(b) Energy scale $T_\text{RKKY}$ rescaled by the square of the coupling $U/\Gamma_0$ as function of the coupling in 1d.
The different lines indicate different values for
$\frac{d}{d\w}\Gamma^-_\mu(\w)|_{\w=0}\propto\frac{R_nk_\text{F}\Gamma_0}{\pi D}$, 
which quantify the applicability of the effective Hamiltonian.
NRG parameters: $\text{N}_\text{s}=4000$ and $\Lambda=3$.
}
\label{fig9}

\end{center}
\end{figure}

For $\frac{d}{d\w}\Gamma^-_\mu(\w)|_{\w=0} \ll 1$, 
the  corrections at large distances
are small, and $H_\text{TIAM}=H_\text{TIAM}^++H_\text{imp}^\text{eff}$ is
a good approximation. 
Our analysis, $J^{AF}_{\rm RKKY} \propto [t^{\rm eff}]^2/U$, demonstrates
that the RKKY interaction should be proportional to $1/U$ instead of the $1/U^2$ dependency expected by
a separate two step transformation: (i) a Schrieffer Wolff transformation onto the two impurity Kondo model
and (ii) the perturbative calculation of $J_{\rm RKKY}$ using this two impurity Kondo model.

Fig.\ \ref{fig9}(a) depicts the local entropy of the impurities for a 1d linear dispersion, plotted against the 
dimensionless temperature $t= T \cdot U/(t^\text{eff})^2$ 
for the distances $R_n$ as defined above but a fixed ratio $R_n k_\text{F}\Gamma_0/D$ 
so that always a  AF RKKY interaction is generated.
The different lines represent different coupling strengths $U/\Gamma_0$ in a range of $1<U/\Gamma_0<60$. 
The FP spectra of the NRG level flow distinguishes the two regimes $J_{\rm RKKY} > T_K$
(black line) and $J_{\rm RKKY} < T_K$ (blue lines). In the upper panel of Fig.\ \ref{fig9}(a)
the corrections can be neglected, $Rk_\text{F}\Gamma_0/D\pi= 0.0075$, and the universal
crossover of the entropy proves that $J_{\rm RKKY} \propto (t^{\rm eff})^2/U$.

This simple scaling does not hold for  a larger 
$Rk_\text{F}\Gamma_0/D\pi= 0.05$ as demonstrated in the lower  panel of Fig.\ \ref{fig9}(a).
Since the crossover to a local singlet should still occur at a temperature scale $J_{\rm RKKY}$
the energy curves suggest a modification from the $1/U$ behavior.

In order to shed some light on 
the $U$ dependency of  $J_{\rm RKKY}$, we
calculated the crossover temperature $T_\text{RKKY}$  as a function of $U/\Gamma_0$.
$T_\text{RKKY}$  is defined as the temperature where the Entropy $S_\text{imp}$ 
has reached the value $S_\text{imp}(T_\text{RKKY})=\frac{1}{2}\ln(4)=\ln(2)$.
Fig.\ \ref{fig9}(b) shows $T_\text{RKKY}\cdot (U\Gamma_0)^2$ 
as function of the coupling strength $U/\Gamma_0$ for different values of $\frac{R_nk_\text{F}\Gamma_0}{\pi D}$.
The  linear increase of the curves for small $U$ proves the $1/U$ dependency. 
For very large $U$, the curves approach a constant. In this regime, $J_{\rm RKKY}\propto 1/U^2$
in accordance with the Schrieffer Wolff transformation onto the TIKM.
The crossover from a charge fluctuation driven $J_{\rm RKKY}\propto 1/U$
to a Kondo interaction driven  $J_{\rm RKKY}\propto J^2 \propto 1/U^2$ does
not only depend on $U$ but is also strongly influenced by ratio $Rk_\text{F}\Gamma_0/D\pi$.
Consequently, the replacement of the TIAM by the TIKM is distance dependent and requires more
care than just investigating the local regimes.

\begin{figure}[t]
\begin{center}
\includegraphics[width=0.5\textwidth]{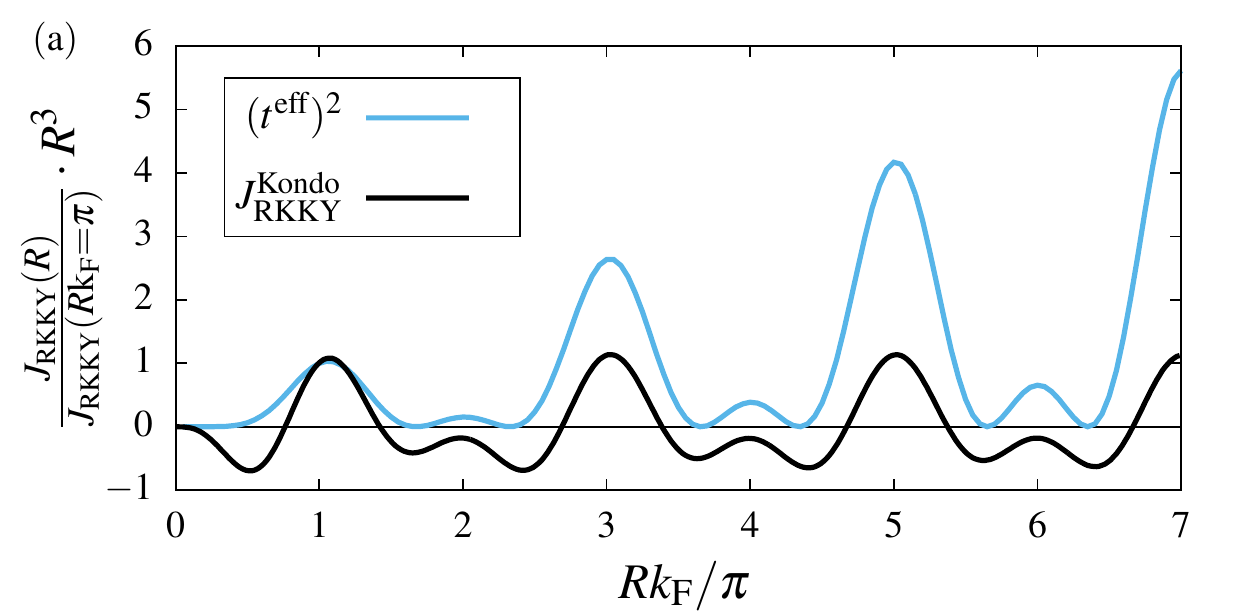}
\includegraphics[width=0.5\textwidth]{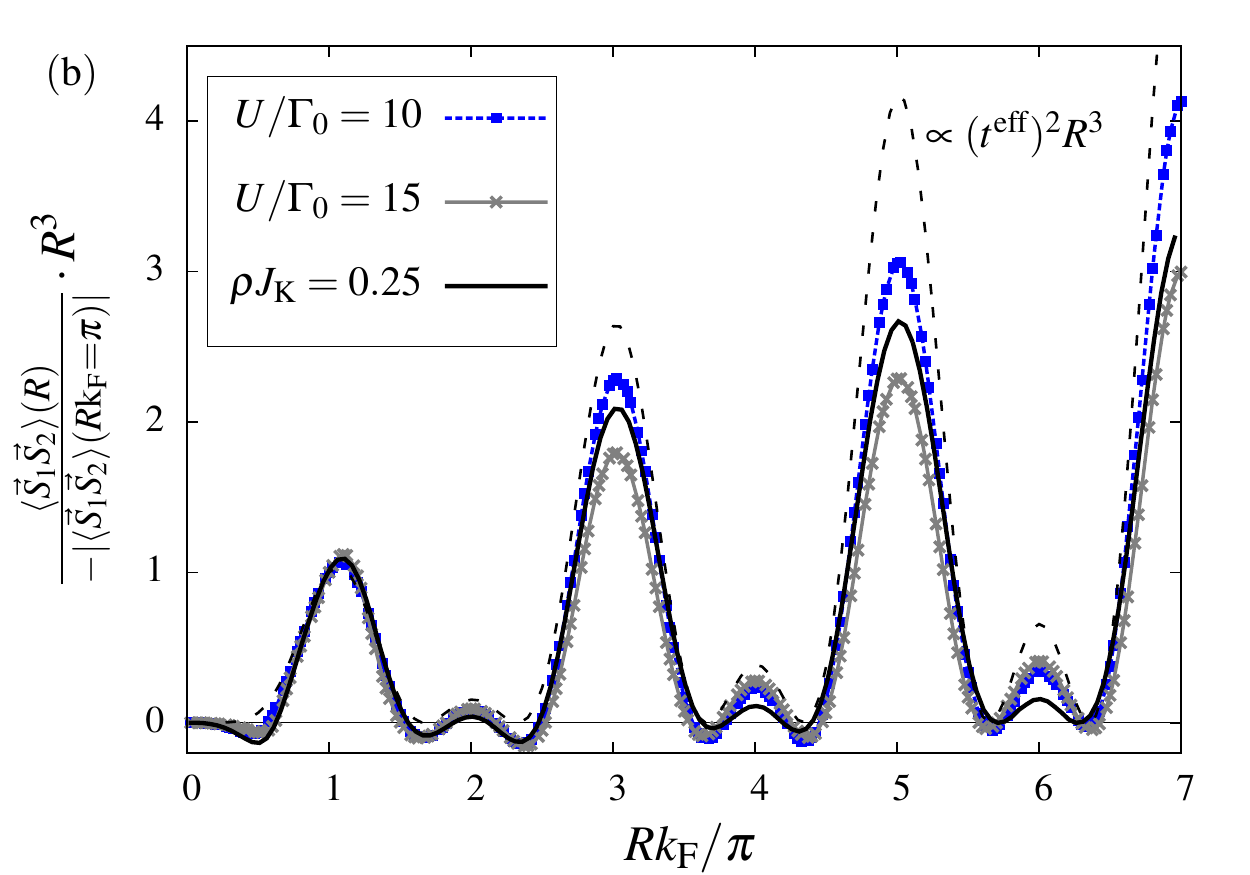}
\caption{
(a) Comparison of the standard RKKY interaction for the TIKM and the
square of the effective tunneling as function of the distance for a 3d linear dispersion.
(b) Impurity spin-spin correlation for the TIKM (black solid line) and the TIAM (lines with points)
as function of the distance for a 3d linear dispersion.
Parameters: $D/\Gamma_0=10,\, \text{N}_\text{s}=4000,\, \Lambda=2$.
}
\label{fig10}

\end{center}
\end{figure}

In case of a linear dispersion in 3d, the corrections \eqref{corrections-3d} are $R$-independent
and the amplitude of $t^{\rm eff}$ always decays as function of $R$. 
Fig.\ \ref{fig10} (a) shows a comparison of   $J^{\rm Kondo}_{\rm RKKY}$ 
calculated by the textbook expression which can be found in
the appendix A of Ref.~\cite{Lechtenberg2014} 
as black line
with $(t^\text{eff})^2$ as light blue line. 
While the envelope function of 
$J_{\rm RKKY}^{\text{Kondo}}$ decays as  $R^{-3}$ as expected from the analytical formula,
$(t^\text{eff})^2\times R^3$ is increasing with distance for small $R$. 
Consequently, $J^{AF}_{\rm RKKY} \propto (t^\text{eff})^2$ decays
as $R^{-2}$ 
in the wide band limit for $U/D \ll1$ in contrary to the expected $R$ dependency of  $J^{\rm Kondo}_{\rm RKKY}$.

While $J_{\rm RKKY}$ describes an effective spin-spin interaction in an effective local moment Hamiltonian,
the $R$ dependency of the spin-spin correlation function is a different property that is governed by the competition 
between the Kondo screening and the RKKY interaction. Fig.\ \ref{fig10}(b) depicts $\langle\vec{S}_1 \vec{S}_2\rangle R^3$
in the TIAM for moderate values of $U/\Gamma_0$ (blue and grey curve with points).
For a better comparison of the decay of the envelope function, we normalized $\langle\vec{S}_1 \vec{S}_2\rangle(R)$ at $Rk_{\rm F}=\pi$,
where the correlations are AF, thus positive values belong to AF correlations.

The $R$ dependence of $t^\text{eff}$
governs the physics in the wide-band limit and for small $U/\Gamma_0$ (dashed bline in Fig.\ \ref{fig10}(b)).
For $U\rightarrow 0$ the analytic equation \eqref{eqn:spin-corr_RML}
proves that the spin-spin correlation function is 
purely AF, whereas the sign of
$J^{\rm Kondo}_{\rm RKKY}$ always oscillates with the distance.
With increasing $U/\Gamma_0$, the FM correlations
emerge continuously from the purely AF function and the
power-law decay of the correlation function seems to cross from those of $(t^\text{eff})^2$ over to  
those of $J^{\rm Kondo}_{\rm RKKY}$ for $U/\Gamma_0\rightarrow \infty$.
Note that we can not resolve this weak coupling regime using the NRG, 
since the numerical noise is  rapidly amplified by the $R^3$ scaling
for $U/\Gamma_0>15$.  
We added  $\langle\vec{S}_1 \vec{S}_2\rangle$ calculated for the two-impurity Kondo model
with a Kondo coupling $\rho J_\text{K}= 0.25$ as a solid black line in Fig.\ \ref{fig10}(b).
Just like small $U/\Gamma_0$ in the Anderson model, large Kondo couplings such as $\rho J_\text{K}= 0.25$ lead to a supression of the
FM correlations due to the Kondo effect \cite{Lechtenberg2018} and a slower decay as $J^{\rm Kondo}_{\rm RKKY}$,
at least in the small distance regime.

%----------------------------------
\section{Simple cubic lattice}
\label{sec:Sclattice}
%----------------------------------
The RKKY interaction has been investigated for more than 60 years and 
it is well established, that the anisotropy caused by
the lattice of the host has a strong influence on the
RKKY iteraction \cite{Lichtenstein2000,Zhou2010,Allerdt2015,Aristov1997,Masrour2016,Nefedev2014}.
However, for a large Kondo-coupling $J_{\rm K}$ and small $U/\Gamma_0$ respectively,
the Kondo effect has a strong influence on the spin-spin correlation function and
the textbook expression for the RKKY interaction is not sufficient to describe the magnetic order \cite{Lechtenberg2018}.
Therefore, the effective tunneling 
for an non spherical band dispersion $\e_{\k}$
provides additional insight to established knowledge
on the RKKY interaction.

In this section, we exemplifies this by focusing on the well studied
simple cubic lattice with lattice spacing $a$ at half band filling.
The dispersion $\epsilon_{\vec{k}}$ in $d$ dimensions takes the form

\begin{eqnarray}
 \epsilon_{\vec{k}}=-\frac{D}{d}\sum_{\alpha=1}^{d}\cos(k_{\alpha}a).
\label{eqn:TB-dispersion}
\end{eqnarray}
Defining a nesting wave vector $\vec{Q}$ and the reciprocal lattice vectors $\vec{G}_\alpha$
\begin{align}
\vec{Q}=\frac{\pi}{a}\sum_{\alpha=1}^{d}\vec{e}_\alpha,\quad \vec{G}_\alpha=\frac{2\pi}{a}\vec{e}_\alpha,
\end{align}
which satisfy the relations $ \epsilon_{\vec{k}\pm\vec{Q}}=-\epsilon_{\vec{k}}$ and $\epsilon_{\vec{k}\pm\vec{G}_\alpha}=\epsilon_{\vec{k}}$, 
we can always find a bijection $f:1.\text{Bz}.\rightarrow 1.\text{Bz}.,\,\vec{k}\rightarrow\vec{k}^\prime$, for which $\epsilon_{\vec{k}^\prime}=-\epsilon_{\vec{k}}$
\begin{align}
 f(\vec{k})=
 \vec{k}^\prime=\vec{k}+\vec{Q}+\sum_{\alpha=1}^d z_{\vec{k},\alpha} \vec{G}_\alpha,\quad z_{\vec{k},\alpha}\in\{\pm1,0\}.
\end{align}
Using this mapping, we analyze the effective densities of states 
with respect to inversion symmetry in energy space
as well as P-H symmetry:
\begin{align}
\Gamma_e(-\epsilon,\vec{R})&=\pi V^2\sum_{\vec{k}^\prime} \delta(\epsilon-\epsilon_{\vec{k}^\prime})\, \text{cos}^2\left\{\vec{k}^\prime\vec{R}/2-\Phi/2\right\},\nonumber\\
\Gamma_o(-\epsilon,\vec{R})&=\pi V^2\sum_{\vec{k}^\prime} \delta(\epsilon-\epsilon_{\vec{k}^\prime})\, \text{sin}^2\left\{\vec{k}^\prime\vec{R}/2-\Phi/2\right\}.
\end{align}
Due to the additional phase $\Phi/2$, 
\begin{align}
\Phi=(\vec{Q}+\sum_{\alpha=1}^d z_{\vec{k},\alpha} \vec{G}_\alpha)\vec{R},
\end{align}
the hybridization function is
P-H symmetric ($\phi=n\pi$) for $R_\alpha/a\in\mathbb{Z}$ only. We can distinguish between the
two different types of symmetries in the following way \cite{Affleck1995}:
\begin{align}
 &\sum_{\alpha=1}^{d} R_{\alpha}=2na\quad&&\longrightarrow\quad \text{first type}
\label{eqn:PH-surface1}\\
 &\sum_{\alpha=1}^{d} R_{\alpha}=(2n+1)a\quad&&\longrightarrow\quad \text{second type}
\label{eqn:PH-surface}
\end{align}
Since the two types of P-H symmetry generate a contribution to the RKKY interaction
with opposite sign,
this result is equivalent to the general RKKY oscillations on a bipartite lattice at half filling \cite{Saeed07}.
Moreover, the effective tunneling vanishes for impurities  placed on the same sub-lattice.

\begin{figure}[t]
\begin{center}
\includegraphics[width=0.5\textwidth]{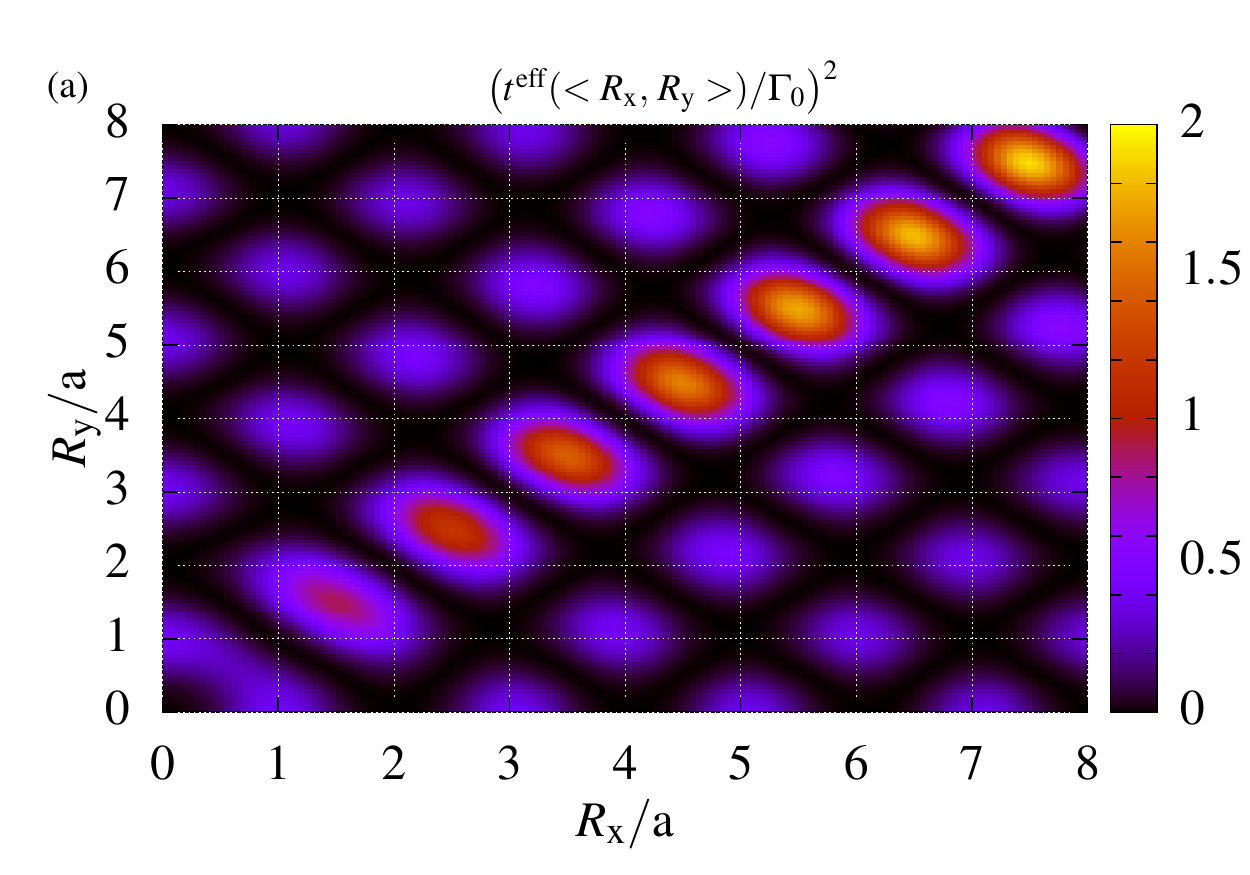}
\includegraphics[width=0.5\textwidth]{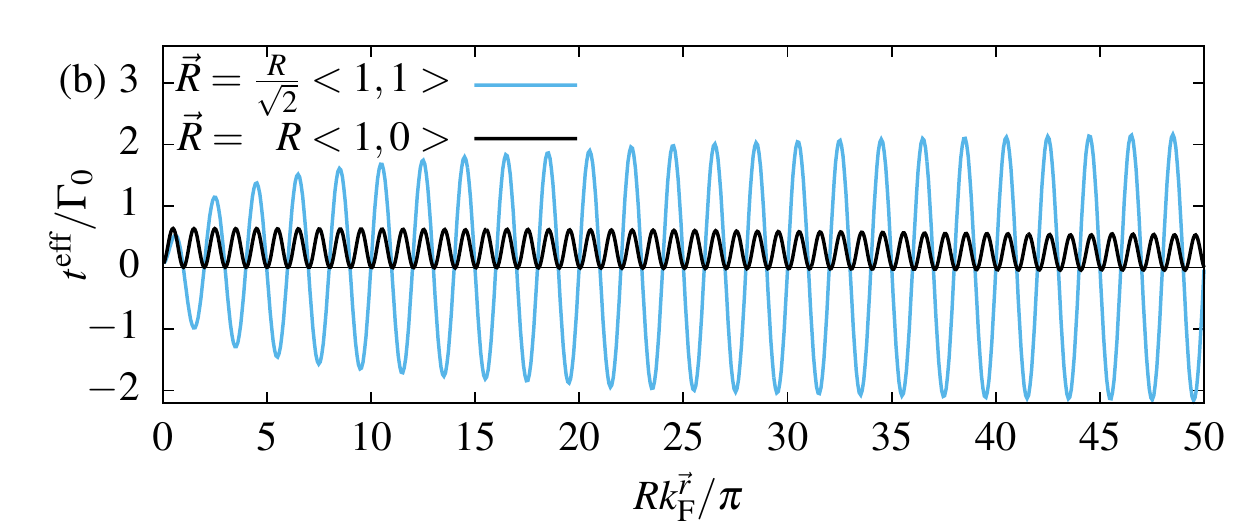}
\caption{
(a) Square of the effective hopping element in two dimensions color-coded as function of the impurity distance $\vec{R}= (R_x,\,R_y)$.
The intersections of the grid correspond to the positions of the host atoms.
%%%
(b) Effective tunneling along the basis and the diagonal direction. The absolute value of the distance is rescaled by $k^{\vec{r}}_\text{F} a /\pi=1/2$
along the basis vector direction and $k^{\vec{r}}_\text{F} a/\pi=\sqrt{2}/2$ along the diagonal direction.
}
\label{fig11}
\end{center}
\end{figure}

\subsection{Two-dimensional lattice at half filling}
\label{sec:2d-half-fill}

Fig.\ \ref{fig11}(a) shows the square of the effective hopping element in two dimensions, color-coded as function of the impurity distance $\vec{R}= (R_x,\,R_y)$.
The periodic structure in the two-dimensional plane indicate 
the importance of well-defined momenta in k-space that govern the RKKY interaction in real space.

In order to gain an analytical insight of the anisotropic structure, 
we rewrite the effective tunneling of Eq. \eqref{eqn:t-eff-analytical} as a sum over the Brillouin zone
\begin{align}
t^\text{eff}(\vec{R})=\gamma\cdot s_\mu\sum_{\vec{k}\not\in \text{FS}}\frac{1+s_\mu\cos(\vec{k}\vec{R})}{\epsilon_{\vec{k}}},
\label{eqn:teff_R}
\end{align}
excluding the Fermi surface (FS) which does not contribute to the principle value integral.
All distance independent constants are merged into the constant $\gamma$.

In a second step, we perform a Fourier transformation into $k$-space
\begin{eqnarray}
t^\text{eff}(\vec{q})&=&
\gamma\int  d^2R \sum_{\vec{k}\not\in \text{FS}}\frac{e^{i\vec{q}\vec{R}}\cos(\vec{k}\vec{R})}{\epsilon_{\vec{k}}}
\nonumber \\
&=&
\left\{\begin{array}{ll} \frac{\tilde{\gamma}}{\epsilon_{\vec{q}}}, & \vec{q}\not\in \text{FS} \\ 0, & \vec{q} \in \text{FS}\end{array}\right.
\label{eqn:teff_q}
\end{eqnarray}
exploiting the fact that  the distance independent part vanishes by symmetry
for a P-H symmetric conduction band. For inversion symmetric dispersion, $\e_{\vec{k}}=\e_{-\vec{k}}$, 
$t^\text{eff}(\vec{q})$ obeys the relation
\begin{eqnarray}
t^\text{eff}(\vec{q})&=& t^\text{eff}(-\vec{q})
\end{eqnarray}
and for P-H symmetry of the conduction band, the condition
\begin{eqnarray}
t^\text{eff}(\vec{R}=0)&=&\int_{-\infty}^\infty t^\text{eff}(\vec{q})d\vec{q}=0
\end{eqnarray}
must hold.

The largest contribution to the $k$-summation in Eq.\ \eqref{eqn:teff_q}
is generated at the anti-nodal points of the dispersion $\e_{\vec{k}}$, 
located at  $\vec{p}_{1/2}=\pm(0,\pi/a)$ and $\vec{l}_{1/2}=\pm(\pi/a,0)$,
and we can approximate the Fourier transformation $t^\text{eff}(\vec{q})$
by a sum of $\delta$-functions
\begin{eqnarray}
t^\text{eff}_\text{anti-nodal}(\vec{q}) 
&\propto&  \sum_{i\in\{1,2\}}\left[\delta(\vec{q}+\vec{p}_i)
-\delta(\vec{q}+\vec{l}_i)\right]
\label{eqn:teff_q_simple}
\end{eqnarray}
with an appropriate prefactor  that is independent of $\vec{q}$.

This simplified expression can be transformed back into real space. 
Along the direction $\vec{n}=\vec{R}/R$,  $R=|\vec{R}|$, 
$t^\text{eff}_\text{anti-nodal}(\vec{R})$ is given by a product of modulations,
\begin{align}
 t^\text{eff}_\text{anti-nodal}(\vec{R})\propto \sin(R k^{\vec{r}}_{\text{F},+})\sin(R k^{\vec{r}}_{\text{F},-})\,.
\label{eqn:teff_frequency}
\end{align}
govern by the two characteristic spatial frequencies 
\begin{eqnarray}
k^{\vec{r}}_{\text{F},\pm}&=&\frac{\pi}{2a} |n_x\pm n_y| \, .
\end{eqnarray}
Along the basis vector direction $\vec{n}=\vec{e}_{\alpha}$, the 
frequencies are identical and the sign of the effective tunneling 
remains positive.

Within this anti-nodal point approximation, the amplitude of the oscillating
effective hopping remains constant. 
This provides a better understanding why full $ t^\text{eff}(\vec{R})$ 
plotted in Fig.\ \ref{fig11}(b) does not decay as function of the distance $R$
and explains the different oscillation frequency in the different spatial directions.

\begin{figure}[t]
\begin{center}
\includegraphics[width=0.5\textwidth]{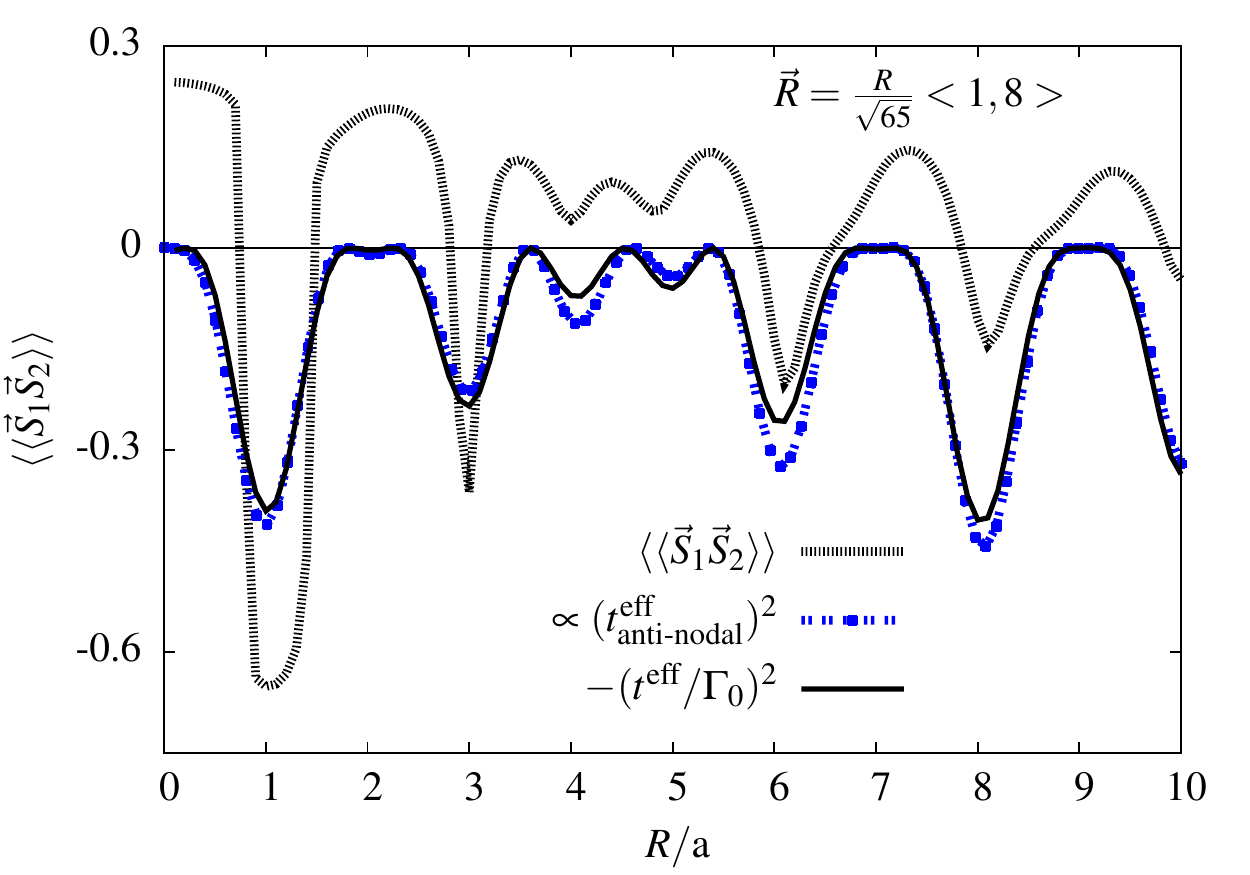}
\caption{Impurity spin-spin correlation function in comparison with the full effective tunneling $t^\text{eff}$ and the contributions
which originate from the anti-nodal points $t^\text{eff}_\text{anti-nodal}$ along the direction $\vec{r}=\frac{1}{\sqrt{65}}(\vec{e}_1+8\vec{e}_2)$ with $D/\Gamma_0=10$ and $U/\Gamma_0=30$.
NRG parameters: $\text{N}_\text{s}=3000,\, \Lambda=4$.
}
\label{fig12}
\end{center}
\end{figure}

To illustrate the quality of the approximation, a comparison between the  full $ t^\text{eff}(\vec{R})$
and   the approximative quantity $t^\text{eff}_\text{anti-nodal}(\vec{R})$ obtained from Eq.\ \eqref{eqn:teff_frequency}, 
is shown for a more generic direction $\vec{n}=\frac{1}{\sqrt{65}}(\vec{e}_1+8\vec{e}_2)$
in Fig.\ \ref{fig12}.
The plot demonstrates that the oscillations of the AF contribution to the RKKY interaction 
are determined by $k^{\vec{r}}_{\text{F},\pm}$.
We also added the spatial dependency of the 
impurity spin-spin correlation function along the same direction as grey dotted curve to Fig.\ \ref{fig12}
illustrating that the sing changes and the oscillatory behavior tracks the spatial dependency of $t^\text{eff}$.

Since $k^{\vec{r}}_{\text{F},-}$, and with that Eq. \eqref{eqn:teff_frequency}, vanishes along the diagonal $r_x=r_y$,
the AF contributions cannot originate from the anti-nodal points $\vec{p}_{1/2}$ and $\vec{l}_{1/2}$ in that case.
Therefore, we reexamine the original expression.
The main contributions to the sum in Eq.\ \eqref{eqn:teff_R} stem from
$k$-points  around the Fermi surface 
\begin{eqnarray}
\vec{k}_\pm^\prime&=&\lim\limits_{\delta \rightarrow 0} \vec{k}_{\in \text{FS}}\pm\delta\vec{n}^\text{FS}_{\vec{k}}
\end{eqnarray}
where $\vec{n}^\text{FS}_{\vec{k}}$ denotes the local normal vector of the FS.
If the oscillations of the numerator $\cos(\vec{k}\vec{R})$ in the vicinity of the FS are small, in general for short distances,
the generic spatial structure should be reproduced by focusing on the summation of a very small $k$-shell
around the FS and we obtain the approximation

\begin{align}
  t^\text{eff}(\vec{R})\approx\tilde{t}^\text{eff}(\vec{R})\propto\sum_{\vec{k}_+^\prime}\frac{\cos(\vec{k}_+^\prime\vec{R})}{\epsilon_{\vec{k}_+^\prime}}+\sum_{\vec{k}_-^\prime}\frac{\cos(\vec{k}_-^\prime\vec{R})}{\epsilon_{\vec{k}_-^\prime}}.
\label{eqn:teff_k-not-FS}
\end{align}
substituting $\epsilon_{\vec{k}_\pm^\prime}=\pm\delta\nabla_{\vec{n}_{\vec{k}}^\text{FS}}\epsilon_{\vec{k}}$
into this expression, we can restrict Eq.\,\eqref{eqn:teff_k-not-FS} 
to a summation over the Fermi surface and a directional derivation along $\vec{n}^\text{FS}_{\vec{k}}$:
\begin{align}
\tilde{t}^\text{eff}(\vec{R})&\propto\sum_{\vec{k}\in\text{FS}}\lim\limits_{\delta \rightarrow 0}\frac{\cos((\vec{k}+\delta\vec{n}^\text{FS}_{\vec{k}})\vec{R})-\cos((\vec{k}-\delta\vec{n}^\text{FS}_{\vec{k}})\vec{R})}{\delta\nabla_{\vec{n}^\text{FS}_{\vec{k}}}\epsilon_{\vec{k}}}\nonumber\\
&=\sum_{\vec{k}\in\text{FS}}\frac{\nabla_{\vec{n}^\text{FS}_{\vec{k}}}\cos(\vec{k}\vec{R})}{\nabla_{\vec{n}^\text{FS}_{\vec{k}}}\epsilon_{\vec{k}}}\,.
\label{eq:74}
\end{align}

The 2d Fermi surface of a simple cubic lattice is given by a square in the first Brillouin zone that is parametrized
by the four conditions $k_x\pm k_y= \pm \pi/a$ and the corners are given by the four anti-nodal points introduced before.
We have shown above that these corners, at which the nominator and denominator in Eq.\ \eqref{eq:74} vanishes, does not
contribute to ${t}^\text{eff}(\vec{R})$ along the diagonal in real space. Therefore we focus on the four nodal points
between the corners given by $(\pm\pi/2a,\pm\pi/2a)$ on the 2d FS. 
The dispersion is linear around these points close to the FS
and $\nabla_{\vec{n}^\text{FS}_{\vec{k}}}\epsilon_{\vec{k}}\approx 2$ in appropriate units.
Therefore, we replace the denominator in  \eqref{eq:74}  by a constant
and integrate over some part of the FS around these nodal points, which can be easily parameterized by a 1d integral,

\begin{align}
  \tilde{t}^\text{eff}_\text{nodal}\left(\vec{R}\right)\propto\left(\int_{\frac{\pi}{2a}-\tau}^{\frac{\pi}{2a}+\tau}dk_x+\int_{-\frac{\pi}{2a}-\tau}^{-\frac{\pi}{2a}+\tau}dk_x\right)\nonumber\\
\biggl[ \bigl(\vec{n}^\text{FS}_{k_x}\vec{R}\bigr)
\sin\Bigl\{\bigl(R_xk_x-R_y|k_x|\bigr)+R_y\frac{\pi}{a}\Bigr\} \biggr].
\end{align}
The distance from the nodal points include here is parametrized by $\tau$, and the 
explicit shape of the FS, $|k_y|=\pi/a-|k_x|$,  was inserted.

\begin{figure}[t]
\begin{center}
\includegraphics[width=0.5\textwidth]{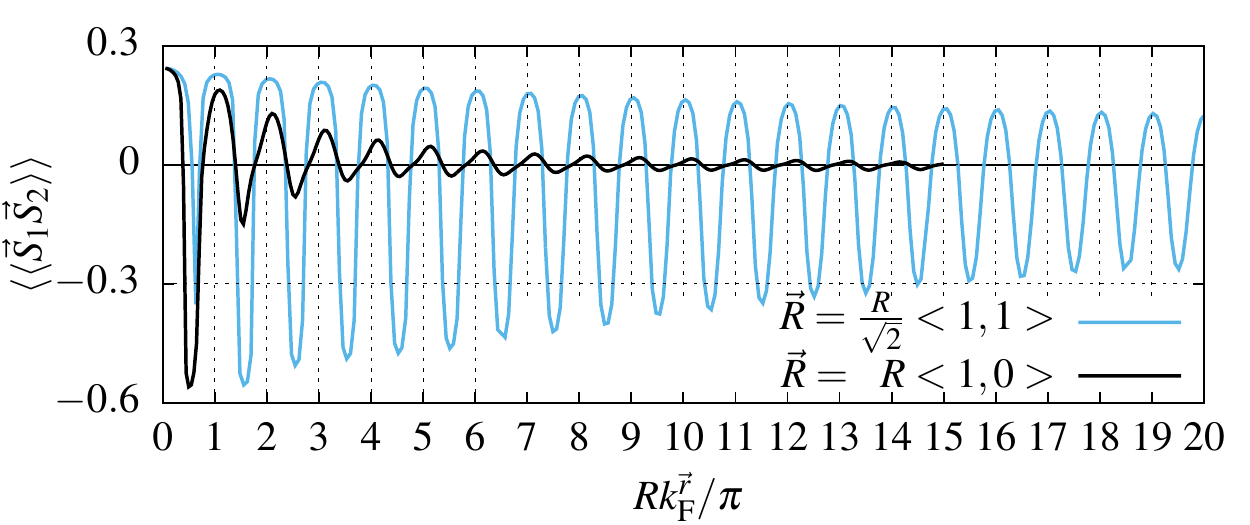}
\caption{
NRG calculations of the impurity spin-spin correlation function along the basis vector and the diagonal direction with $D/\Gamma_0=10$ and $U/\Gamma_0=30$.
NRG parameters: $\text{N}_\text{s}=3000,\, \Lambda=4$.
}
\label{fig13}
\end{center}
\end{figure}

For a general direction, this leads to small contributions due to the oscillations of the integrand.
Only along the diagonal direction, 
these oscillations cancel. In addition $R_y\pi/a=Rk^{\vec{r}}_{\text{F},+}$ holds
and we  find a linear increase  with the distance $R$
\begin{align}
  \tilde{t}^\text{eff}_\text{nodal}\left(\frac{R}{\sqrt{2}}\begin{pmatrix}1\\1\end{pmatrix}\right)\propto R\,\sin\left(Rk^{\vec{r}}_{\text{F},+}\right).
\label{eqn:teff_diagonal}
\end{align}
This provides the deeper understanding of the surprising increase of the amplitude of the full $t^\text{eff}(\vec{R})$
shown in Fig.\ \ref{fig11}(b). Our analytical calculation links this observation to the properties of the dispersion
at the nodal points of the FS for the P-H symmetric band.
Note that for large values of $R$, the oscillations around the FS
in the nominator of Eq.\, \eqref{eqn:teff_k-not-FS} lead to a damping of the linear increase:
The simplification entering Eq.\ \eqref{eq:74} are not valid for large $R$

For larger distances $R$, the amplitude of the tunneling stays constant in all directions
as a consequence of the perfect FS-nesting for a 2d simple-cubic dispersion at half-filling.
The linear increase of the effective tunneling along the diagonal direction
strongly depends on the structure of the FS, but is not a consequence of FS-nesting and the divergency of the 
Lindhard function in momentum space respectively. 

The spatial and band width 
corrections to the spin correlation function discussed in Sec.\  \ref{sec:RKKY} predict a 
decay of the correlation function even for
constant $t^{\rm eff}$ that contains the exact AFM RKKY interaction in the limit $\Gamma_0/D\rightarrow0$. 
Fig.\ \ref{fig13} depicts the impurity spin-spin correlation as function of the distance along the basis and the diagonal direction.
As expected from the initially linearly increasing and than constant
$t^{\rm eff}$ in the diagonal direction, the correlation function dominates in this direction.
Its amplitude only shows a slow decay as function of $R$ caused by the finite band width.

\subsection{Particle and hole doping in two dimensions} 

The strong influence of the explicit shape of the Fermi surface on the effective tunneling and the RKKY interaction
respectively, can be demonstrated by adding an additional chemical potential $\mu$ which influences the energy dispersion of the host
and the level energy of the impurities
\begin{align}
\epsilon_{\vec{k}}\rightarrow\epsilon_{\vec{k}}+\mu\,;\quad \epsilon^f\rightarrow\epsilon^f+\mu.
\end{align}
Obviously the chemical potential breaks P-H symmetry in the initial conduction band but preserves parity.
Our analysis, however, remains valid and the P-H asymmetry can be still casted
into an distance dependent hopping term
\begin{eqnarray}
t^\text{eff}(\vec{R})&=&
\Re(\Delta^-_e(0,\vec{R}))-\Re(\Delta^-_o(0,\vec{R}))
\nonumber \\
&=& \gamma\cdot \sum_{\vec{k}\not\in \text{FS}}\frac{\cos(\vec{k}\vec{R})}{\epsilon_{\vec{k}}} 
\end{eqnarray}

In 2d  simple cubic lattice with a nearest neighbor tight binding description of  the band dispersion,
the sign of the chemical potential determines the 
topology of the FS: A negative value of $\mu$ leads to 
a spherical structure of the FS, whereas a positive potential 
induces general hole pockets.

\begin{figure}[t]
\begin{center}
\includegraphics[width=0.5\textwidth]{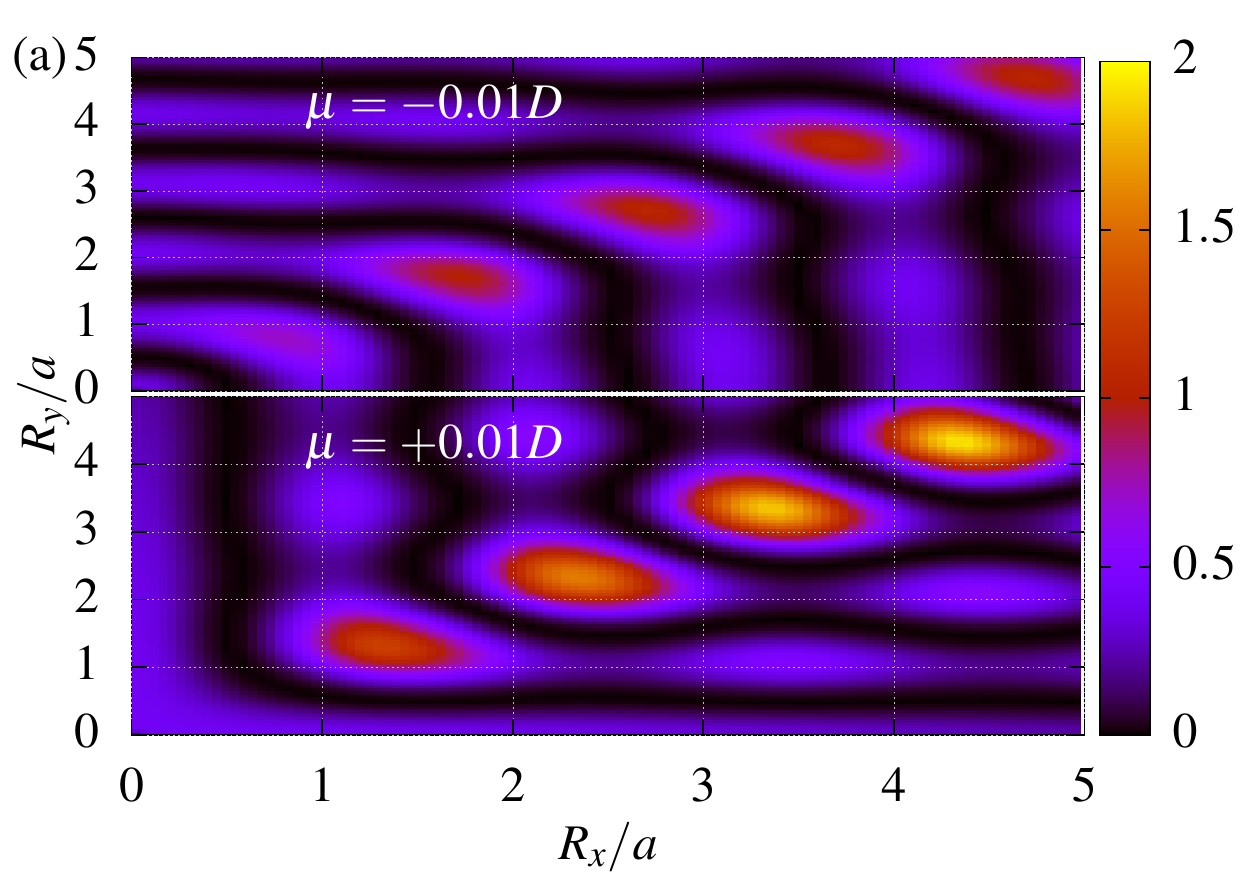}
\includegraphics[width=0.5\textwidth]{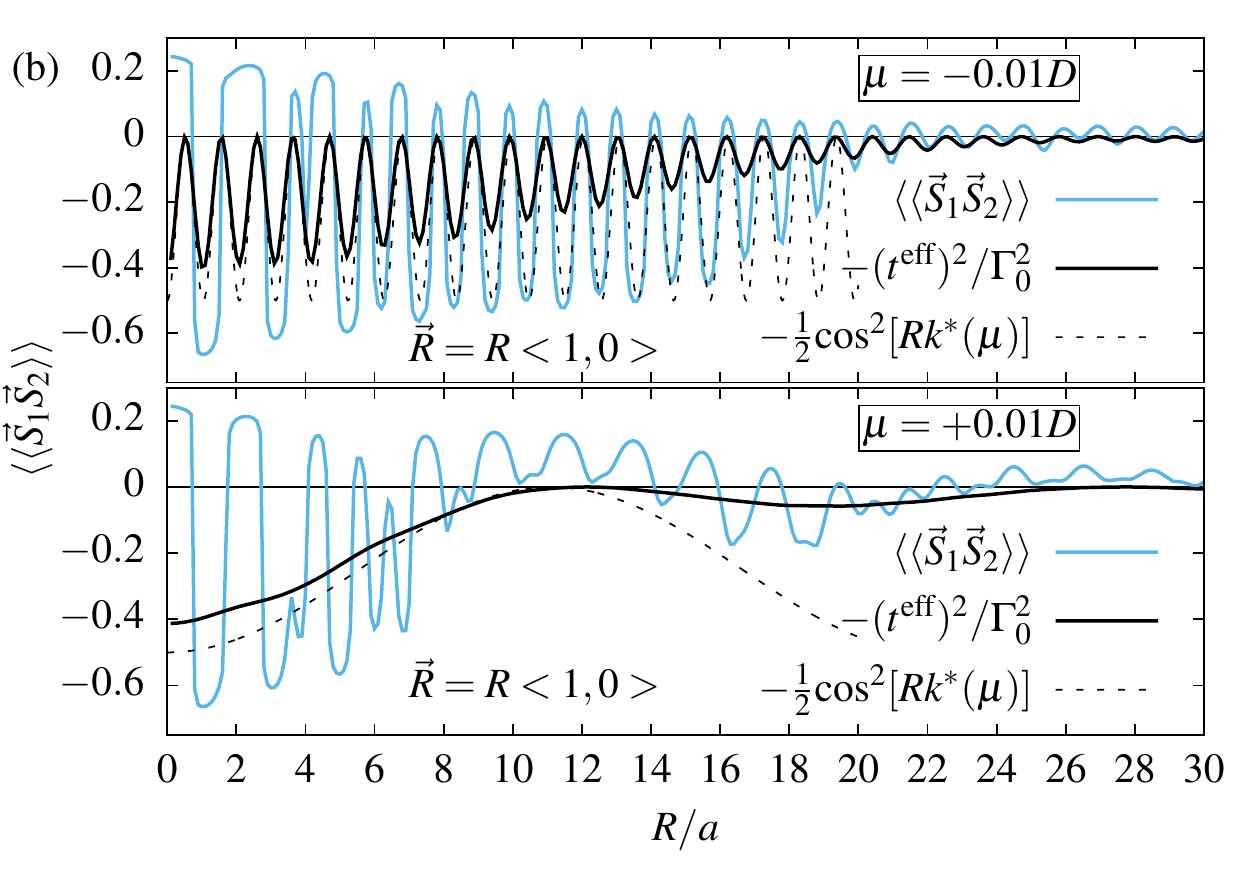}
\caption{(a) Square of the effective hopping element in 2d as function of the impurity distance $\vec{R}=<R_x/a,\,R_y/a>$.
The frequency along the basis vector directions strongly depends on the sign of the chemical potential and the topology of the FS
respectively. The intersections of the grid correspond to the positions of the host atoms.
(b) NRG calculation of the impurity spin-spin correlation function (blue line)
and the square of the effective tunneling (black line)
along the basis vector direction. We added analytically extracted spatial
contribution
$\cos^2(k_x^*(\mu) R)$ as dashed line.
Parameters: $D/\Gamma_0=10$, $U/\Gamma_0=30$, $\text{N}_\text{s}=3000,\, \Lambda=4$.
}
\label{fig14}
\end{center}
\end{figure}

In  section \ref{sec:2d-half-fill}, we have shown that either the anti-nodal or the nodal points 
of the square FS are responsible for
the main contributions to the effective tunneling, depending on the directional alignment of the impurities. 
Since a very weak doping away from half filling only deforms the FS around the anti-nodal points, 
we expect a strong influence 
on the effective tunneling only along the basis vector direction.
Fig.\ \ref{fig14} (a) depicts the color-coded effective tunneling for electron and hole doping,
i.~e.\ holelike and spherical  FS. 
While the general structure along the diagonal direction matches the P-H symmetric case,
the spatial frequency along the basis vector direction varies significantly.

To understand this  change in the frequency in the direction of the basis vectors, we focus on 
$t^\text{eff}_x(R)=t^\text{eff}(R\vec{e}_x)$ and perform a one-dimensional  Fourier transformation 
of \eqref{eqn:teff_R}
\begin{eqnarray}
t^\text{eff}(q)&=&
\gamma\int_{-\infty}^\infty d R\sum_{\vec{k}\not\in \text{FS}}\frac{e^{iq R}\cos(k_x R)}{\epsilon_{\vec{k}}}\nonumber\\
&=& 
\frac{\pi \gamma}{t}\sum_{\vec{k}\not\in \text{FS}}\frac{\delta(q-k_x)  + \delta(q+k_x) }{
\cos(k_x a) + \cos(k_y a) +\mu/t} \, ,
\label{eqn:teff_q_alpha}
\end{eqnarray}
where we substituted $\epsilon_{\vec{k}}$ defined with respect to the chemical potential.

We have to perform a $k_y$ summation for every $q$ while $k_x$ is fixed by $\delta$-functions.  
$t^\text{eff}(q)$ has the largest contribution for those $q$ values for which many k-vectors $(\pm q,k_y)$
are very close to the FS.

We recall that upon hole doping the Fermi-surface shrinks and become more spherical.
While the nodal points remain almost unaltered, the major change occurs in the vicinity of the anti-nodal points which are shifted
to smaller $k_x$  ($k_y$) values for $k_y=0$ ($k_x=0$). For negative $\mu$, the major contribution arises
from the intersection of the FS with the $k_x$ axis since the FS is perpendicular to the axis at theses shifted anti-nodal points.
Solving  $\epsilon_{k_y=0}=0$ for $k_{x}^*(\mu)$ yields
\begin{eqnarray}
k_x^*(\mu)a  &=& \arccos(|\mu|-1)
\end{eqnarray}
and therefore, the major contributions stem from large $k_x^*(\mu)$ that develop adiabatically from $k_x=\pi$. 
Simultaneously, the contributions from the second pair of anti-nodal points, $(0,\pm \pi/a)$ rapidly vanishes with increasing
hole doping. At the end, we are left with 
\begin{align}
 & t^\text{eff}(q,\mu)\approx\delta(q+k^*_x(\mu))+\delta(q-k^*_{x}(\mu))\\
\Leftrightarrow\quad & t^\text{eff}(R,\mu)\propto \cos[Rk^*_{x}(\mu)]\,.
\end{align}
The missing contribution for $k_x=0$ leads to a doubling of the spatial frequency  away from half filling as
can be seen in Fig.\ \ref{fig14}(a) along the $x$ ($y$)-axis  compared to Fig.\ \ref{fig11}(a).

The situation is qualitatively different for electron doping ($\mu>0$), where the FS is formed by the four hole pockets.
The FS does not intersect with either $k$-axis. However, the FS becomes parallel to the $k_y$ axis close
to the Brillouin zone boundary for a small value $k_{x}^*(\mu)$,
\begin{eqnarray}
k_{x}^*(\mu) &=& \arccos(1-\mu) \, .
\end{eqnarray}
$k_{x}^*(\mu)$ evolves from $k_x=0$ at half-filling, while the contribution from $k_x=\pi /a$ vanished rapidly with increasing $\mu$.
As a consequence the spatial oscillation of  $t^\text{eff}(\vec{R})$ along the $x$ or $y$ axis are very slow as shown
in Fig.\ \ref{fig14}(a) for $\mu=0.01D$.

A spherical deformation of the FS leads to a fast oscillation of the effective tunneling 
with $k^*_{x}(\mu)\approx2\cdot k^{\vec{r}}_{\text{F},\pm}$,
in contrast to slow oscillations in the presence of hole pockets. 
This is illustrated for the two different cases in Fig.\ \ref{fig14}(b)
where the full $t^\text{eff}(\vec{R})$ (black solid curve) is compared to 
the main contribution stemming from the spatial frequency $k_{x}^*(\mu)$(dashed line).

\begin{figure}[t]
\begin{center}
\includegraphics[width=0.5\textwidth]{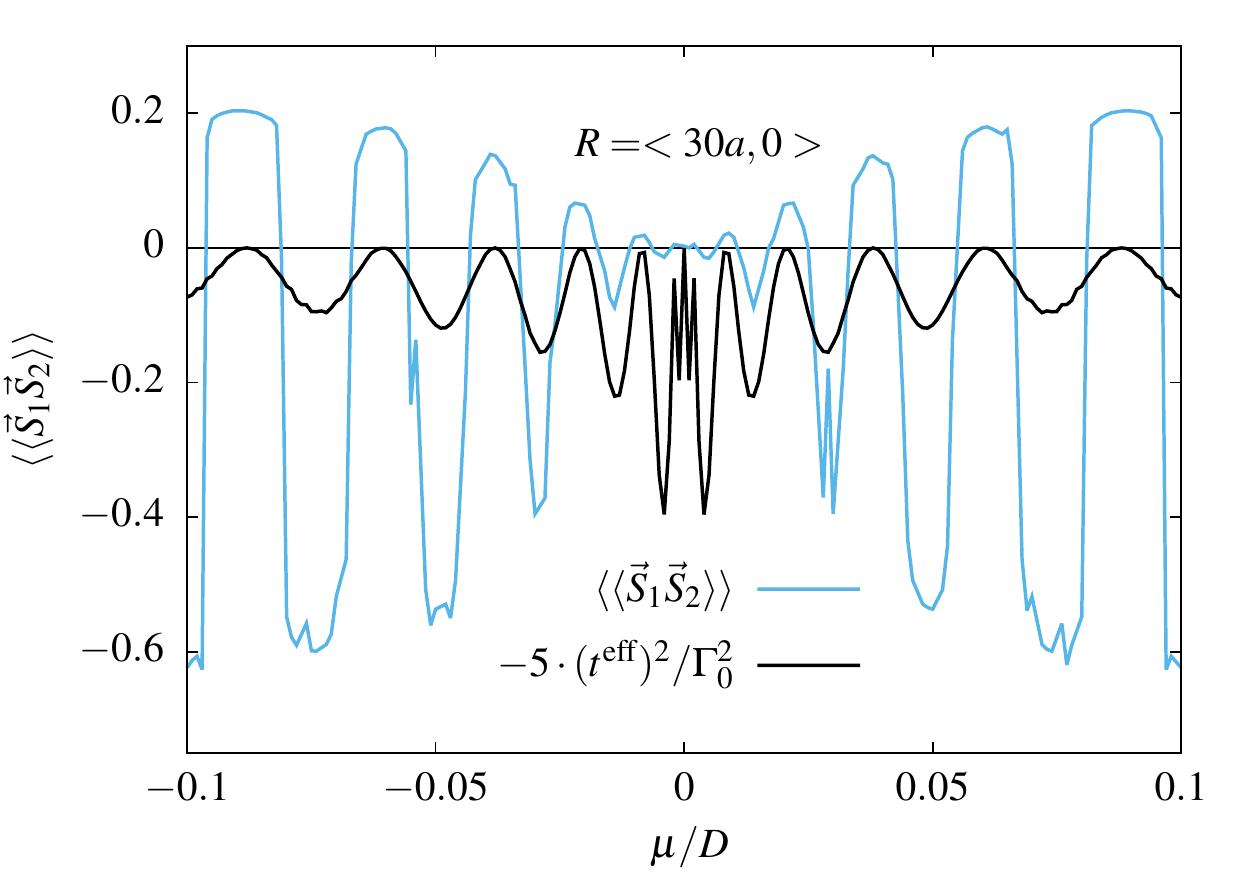}

\caption{NRG calculation of the impurity spin-spin correlation function for a coupling of $U/\Gamma_0=30$ and the square of the effective tunneling, plottet against
the dimensionless chemical potential. The RKKY interaction oscillates as function of $\mu/D$.
Parameters: $D/\Gamma_0=10$, $U/\Gamma_0=30$, $\text{N}_\text{s}=3000,\, \Lambda=4$.
}

\label{fig15}
\end{center}
\end{figure}

We augment this analysis for  $t^\text{eff}(\vec{R})$
with the  NRG calculation of the  impurity spin-spin correlation function
along the basis vector direction, for positive and negative values of the chemical potential.

For a negative chemical potential and a spherical FS, 
the antiferromagnetic part of the correlation function almost shows the same oscillations
as the square of the effective tunneling as can be seen in the upper panel of Fig.\ \ref{fig14}(b).
The small deviations of the correlation function from the behavior of the effective tunneling can be ascribed to the FM part of the RKKY interacting
which is not captured by the effective tunneling and evolves for finite $U/\Gamma_0$.

In the presence of hole pockets, lower panel of Fig.\ \ref{fig14}(b),
the general characteristics of the slow oscillations can be identified in the impurity spin-spin correlation function, too.
In the vicinity of a vanishing effective tunneling, only ferromagnetic correlations 
are observed. The sign of the  correlation function oscillates only in the presence
of an antiferromagnetic contribution to the RKKY interaction, generated by $t^\text{eff}(\vec{R})$.

Fig. \ref{fig15} shows the spin-spin correlation function (blue curve) as well as the $[t^\text{eff}]^2$
(black curve) as function of $\mu$ for a constant impurity distance $\vec{R}$
in order to illustrate the non-linear dependence of the frequency of the spin-spin correlation function
on the chemical potential.

\subsection{Three-dimensional lattice at half filling}

In the previous sections, we demonstrated that the richer spatial dependency of the spin-spin
correlation function as well as the effective tunneling, beyond the simplified isotropic 
$2k_F$ oscillations, originates from the generically non-spherical FS and can be analysed by the investigation of
the analytical properties of the integrals.

\begin{figure}[t]
\begin{center}
\includegraphics[width=0.5\textwidth]{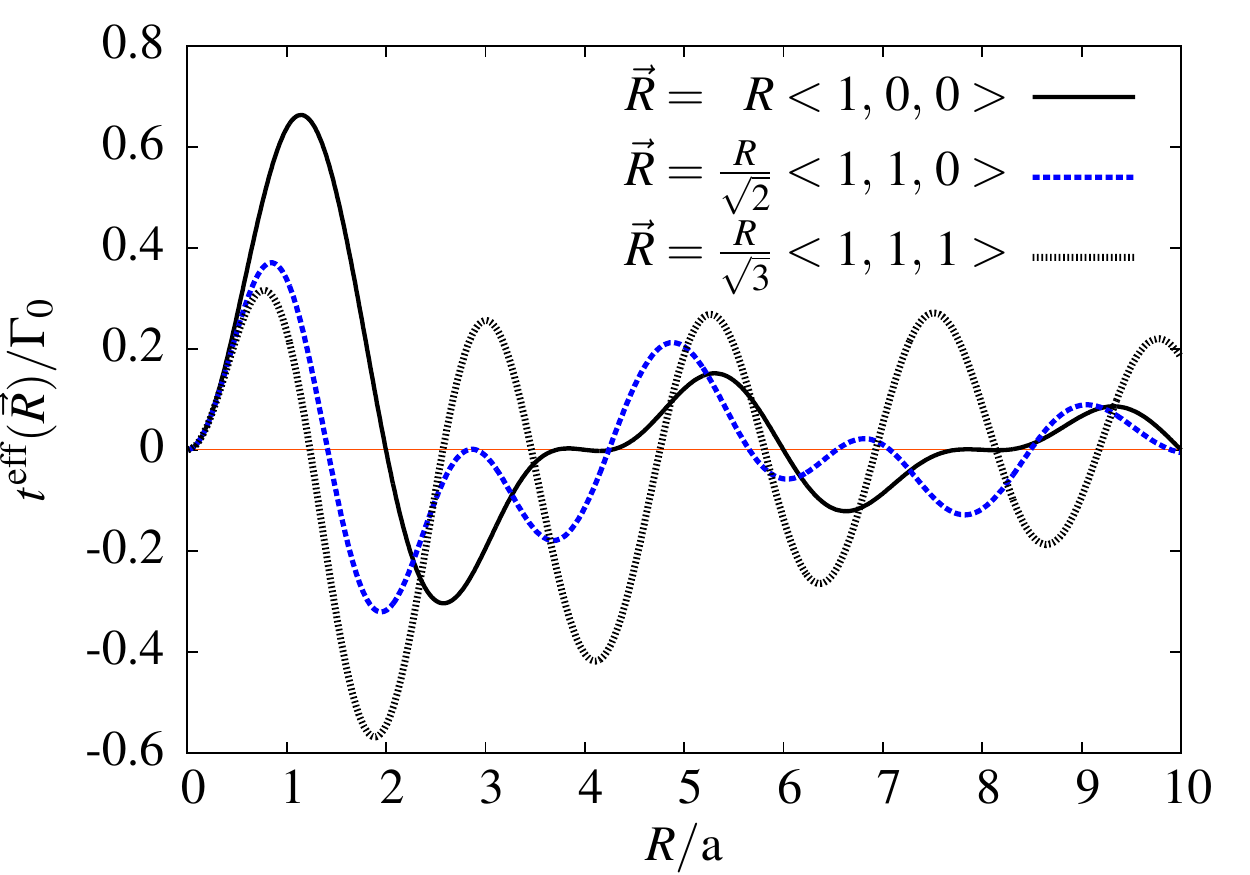}

\caption{Effective hopping element in 3d plotted against the dimensionless impurity distance $R/a$. The anisotropic lattice structure is mirrored
in a strong dependence of the spatial frequency on the directionality. 
}
\label{fig16}
\end{center}
\end{figure}

We now extend our study to the 3d simple cubic dispersion.
The effective tunneling term  
along the 
three symmetry directions is depicted in Fig.\ \ref{fig16}.
Just like in two dimensions, the superposition of different frequencies account for complex oscillations.
The symmetry properties on the lattice places, defined by Eq.\ \eqref{eqn:PH-surface1} and \eqref{eqn:PH-surface},
are fulfilled.

\subsubsection{AFM RKKY coupling - The Doniach scenario}

In section \ref{sec:restoring_the_QCP} 
we discussed the transition from the Kondo singlet to an inter-impurity singlet groundstate
as function of an external applied magnetic exchange-interaction $J_{12}$. In principle, the transition can also
be realized by changing the ratio between the antiferromagnetic RKKY interaction and the Kondo temperature. 
The transition is typically 
discussed in terms of the Kondo coupling $J_\text{K}$ in the context of  the TIKM.
In lattice systems this is referred to  as 
the Doniach scenario \cite{Doniach1977}: the heavy fermi liquid \cite{Grewe91}
is replaced at the QCP by
an AF ordered state generated by the inter-impurity singlets in a lattice.

Decreasing $J_\text{K}$ causes an exponential
decay of $T_\text{K}$, whereas the RKKY interaction only falls of as $J_\text{K}^2$
leading to a increase of the ratio $J_\text{RKKY}/T_\text{K}$.
In the TIAM, the Kondo coupling $J_\text{K}$ is related to the ratio of the Coulomb interaction $U$
and the coupling strength $\Gamma_0$.

Linneweber et al.\ investigated the TIAM on the three-dimensional simple cubic lattice 
via a Gutzwiller variational approach and found such a QCP at a critical $U_c$
provided the impurities are placed on the lattice sites \cite{Linneweber2017}.
The authors indicated that their QCP is probably an artifact of the Gutzwiller variational approach:
the Gutzwiller trial wave function  only includes local correlations on the impurity site while
the NRG reveals already for the Kondo problem the extended nature of the correlated singlet 
\cite{Borda2007,Lechtenberg2014}.

\begin{figure}[t]
\begin{center}
\includegraphics[width=0.5\textwidth]{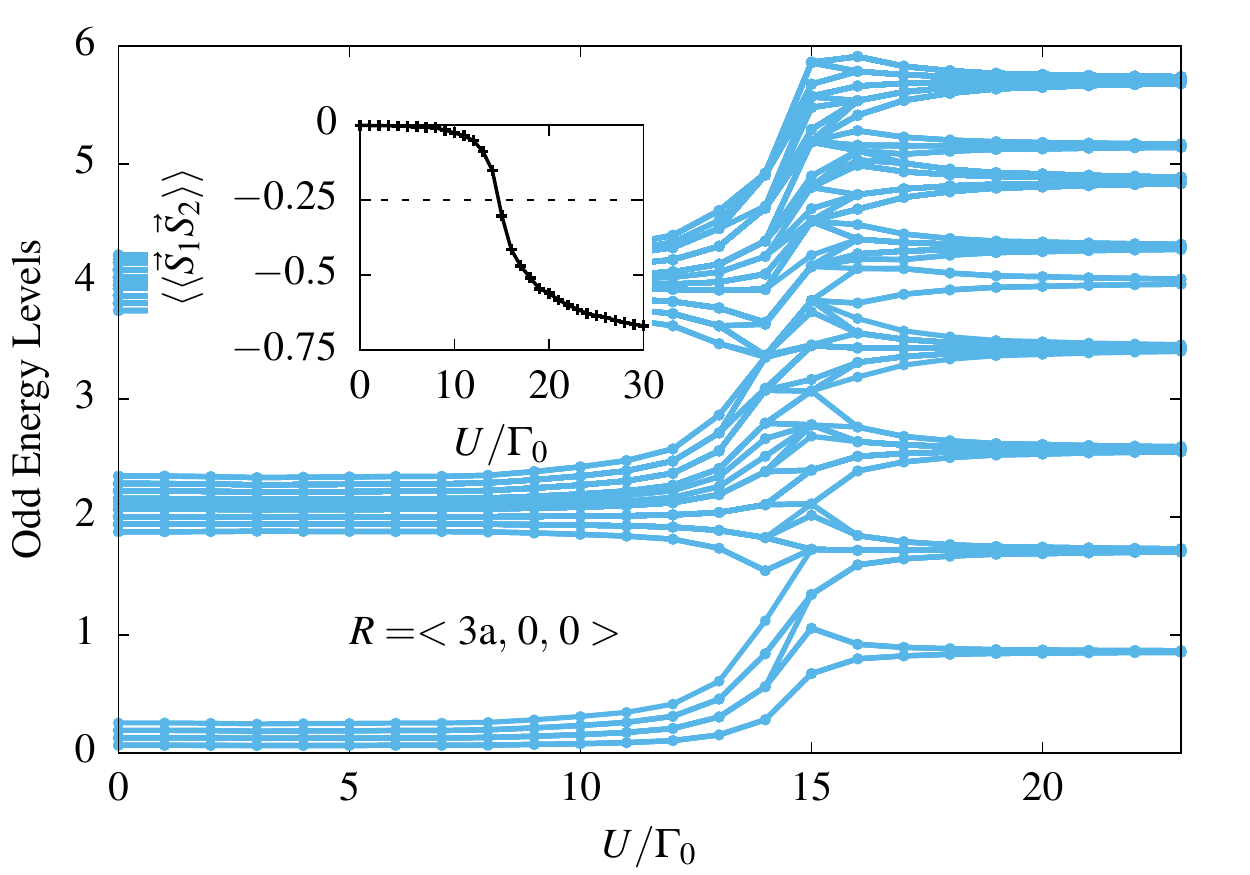}

\caption{Doniach scenario in the TIAM for an impurity distance $\vec{R}=<3a,0,0>$. The inset depicts the 
impurity spin-spin correlation function and illustrates the interimpurity singlet formation for large values of $U/\Gamma_0$.
The continous energy flow between the two fixed points displays a crossover.
NRG parameters: $\text{N}_\text{s}=3000,\, \Lambda=4$.
}
\label{fig17}
\end{center}
\end{figure}

If the impurities are placed on different sublattices, e.g. if they are separated by an odd number of the
lattice spacing along the basis vector direction $\vec{R}=R^\text{odd}<1,0,0>$, 
the RKKY interaction is always antiferromagnetic.
The NRG level flow of the stable FP as function of $U/\Gamma_0$ is shown in Fig. \ref{fig17}.
Clearly, the FP changes continuously from the SC fixed point with P-H symmetry breaking scattering term
to the inter-impurity singlet FP with the absence of a phase shift of the conduction electron states.
The crossover occurs in the vicinity of $U/\Gamma_0\approx 14$.
The inset of Fig.\ \ref{fig17} depicts the corresponding impurity spin-spin correlation and illustrated
the formation of a local inter-impurity singlet in the limit of $U/\Gamma_0\rightarrow \infty$. 
Non indication of a QCP is found  by the NRG when increasing $U$.

The absence of the QCP originates from the fact, that the RKKY interaction cannot only be reduced to an spin exchange interaction.
The RKKY driven charge exchange between the impurities, which is responsible for the antiferromagnetic interaction,
generates marginal relevant operators in the renormalization flow, driving the system away from the QCP.

In order to restore the QCP as function of $U/\Gamma_0$,
a compensating  effective tunneling $-t^\text{eff}(\vec{R})$
has to be added as well as
an additional, antiferromagnetic spin exchange $J_{12}$
that would control the distance to the Varma-Jones type QCP.

\subsubsection{FM RKKY coupling}

\begin{figure}[t]
\begin{center}
\includegraphics[width=0.5\textwidth]{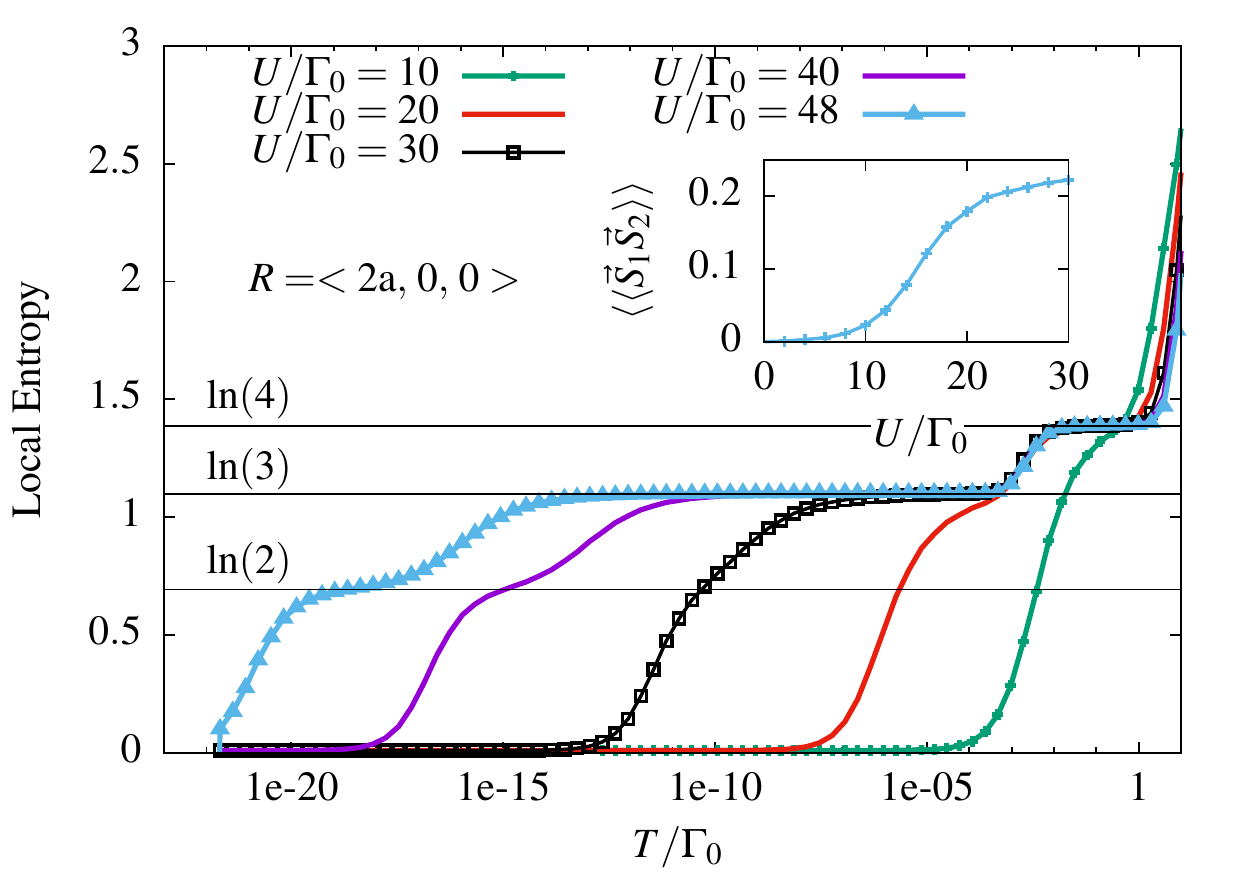}

\caption{
The impurity entropy vs $T$ for different values of $U$ for an impurity distance $\vec{R}=<2a,0,0>$.
Inset: impurity spin-spin correlation function.
NRG parameters: $\text{N}_\text{s}=3000,\, \Lambda=4$.
}
\label{fig18}
\end{center}
\end{figure}

If the impurities are placed on the same sublattice, the RKKY interaction is 
ferromagnetic according to Eq.\ \eqref{eqn:PH-surface1}.
After the local moments are formed, they align with increasing RKKY interaction, and 
the resulting triplet state
is screened in a two-stage Kondo effect \cite{Jones1987}. This is clearly visible by
tracking the impurity entropy as function of temperature \cite{Mitchell2015} for different values of $U$ as depicted
in Fig.\ \ref{fig18}. The local moment formation 
occurs on a scale of $U/\Gamma_0$ leading to a $\ln(4)$ impurity entropy
contribution at intermediate temperatures. By lowering $T$ further,
we observe the crossover to a local triplet  on the scale defined by $J_{\rm RKKY}$:  The larger
$U$ is,  the more pronounced the consecutive two stage Kondo screening is visible revealing the different
unstable FP of the RG flow.

The inset of Fig. \ref{fig18} shows the impurity spin-spin correlation as function of $U/\Gamma_0$ and 
displays the formation of a local inter-impurity triplet. 
Since the Kondo temperature is suppressed with
increasing $U$, the RKKY interaction dominates at higher temperature and favors a correlated triplet
that is collectively screened in a two stage process for $T\to 0$.

We do not find a breakdown of the Kondo effect in the presence of FM RKKY interaction 
found in a recent perturbative RG treatment of the TIKM \cite{Kroha17}.
While this RG approach focuses on the renormalization of the effective Kondo coupling at one
of the impurity sites, the NRG includes all couplings for both impurities on equal footing.

%----------------------------------
\section{Summary and conclusion}
\label{sec:conclusion}
%----------------------------------

Mapping the TIAM  onto degrees of freedom with even and odd parity
symmetry  generates two, in general P-H asymmetric, hybridization functions.
Both hybridization functions can be decomposed into a 
symmetric part with respect to the frequency and an asymmetric
correction. Neglecting the asymmetric part generates always
a FM RKKY interaction. 
A QCP as observed by Varma and Jones is found after adding an direct antiferromagnetic 
exchange interaction. 
The asymmetric part, however, is responsible for an additional relevant
scattering term at zero energy and  hence destroys the Varma and Jones QCP.

We have shown that the  effect of the
asymmetric part  is equivalent to an effective tunneling term
between the two impurities: 
Replacing the full hybridization function by the symmetrized contribution and this local 
tunneling term leads to the identical low-temperature FP  spectrum in the NRG.

This opened the door for restoring the QCP by adding a suitable tunneling term 
to the full Hamiltonian at a fixed distance R or by adjusting the distance between the impurities.
While the counter term can be analytically calculated  for P-H symmetric impurities, 
the term is determined by an numerical iteration procedure for 
P-H asymmetric impurities. 
Using the estimates from the case of P-H symmetric impurities
as the initial value, the parameter $t^{\rm eff}$ or $R$ are iteratively adjusted such that
the lowest single-particle excitation in the even and the odd channel are equal. 
We checked the consistence with the scattering phases of both single-particle Green functions
and found that both phases are also identical at the QCP.

Using the replacement of the full model by a symmetric 
hybridization function and a
local tunneling term, provides an better  understanding of the antiferromagnetic contribution
to the RKKY interaction. 
In contrary to the RKKY interaction of a two-impurity Kondo model resulting from a Schrieffer Wolff transformation,
we find $J_{\rm RKKY}^{AF} \propto (t^{\rm eff})^2/U$.
Only for very large $U$, a crossover to 
$J_{\rm RKKY}^{AF} \propto 1/U^2$ is observed.
Furthermore, the  value of $J_{\rm RKKY}$ decays much more rapidly than
$J_{\rm RKKY}^{AF}$. The distance dependency of the corresponding 
spin-spin correlations, however, tracks the distance dependency of $(t^{\rm eff})^2$
indicating a significant derivation for small and intermediate $U$.

For a constant tunneling $t^{\text{eff}}$, the impurity spin-spin correlation function is governed by a dimensionless
variable that accounts for the distance-dependent correction and the correction to the wide-band limit.

We analyzed the spatial anisotropy of the RKKY interaction as well as the 
impurity spin-spin correlation function for square lattices in different dimensions.
We identified the major  spatial frequencies that 
governs the oscillation in real space for different chemical potentials
close to half-filling and link them to the single-particle dispersion as well as 
the shape of the Fermi surface.

No QPT is found upon increasing the local Coulomb repulsion
for  a distance at which the RKKY interaction is AF.
The spin-spin correlation function  as well as the NRG 
low-temperature FP spectrum changes continuously 
from the strong coupling FP to towards the flow  characterizing the local singlet 
phase.
For a distance leading to a FM RKKY interaction, a two-stage Kondo effect only becomes more pronounce 
with increasing $U$. Therefore, the Doniach scenario for Heavy Fermion QCP requires 
lattice effects that are included in the particle-hole Bethe saltpeter equation 
that enters the lattice spin susceptibility.

\begin{acknowledgments}
We acknowledge fruitful discussions with J.\ B\"unemann, F.\ Gebhard
and H.\ Kroha. 
B.L. thanks the Japan Society for the Promotion of Science (JSPS) and the Alexander von Humboldt Foundation.

\end{acknowledgments}

%%----------------------------------
%\appendix
%%----------------------------------

%----------------------------------

%\bibliography{references}

%merlin.mbs apsrev4-1.bst 2010-07-25 4.21a (PWD, AO, DPC) hacked
%Control: key (0)
%Control: author (8) initials jnrlst
%Control: editor formatted (1) identically to author
%Control: production of article title (-1) disabled
%Control: page (0) single
%Control: year (1) truncated
%Control: production of eprint (0) enabled
%

\end{document}